\documentclass[twocolumn,preprintnumbers,amsmath,amssymb,a4paper ]{revtex4}
\usepackage{fullpage}
\usepackage{amsmath}
\usepackage{amsfonts}
\usepackage{amssymb}
\usepackage{color}
\usepackage{graphicx, graphics}
\usepackage{epsfig}
\usepackage{epstopdf}
\usepackage{subfigure}
\usepackage{wasysym}
\usepackage[version=3]{mhchem} 
\usepackage{sidecap}
\usepackage{changes}
\usepackage{bm}
\usepackage{floatrow}
\usepackage{color}
\usepackage{lipsum}
\usepackage[T1]{fontenc}
\usepackage[final]{pdfpages}
\usepackage{pdfpages}
\usepackage{lipsum}

\makeatletter
\@addtoreset{section}{part}
\def\@part[#1]#2{%
    \ifnum \c@secnumdepth >\m@ne
      \refstepcounter{part}%
      \addcontentsline{toc}{part}{\thepart\hspace{1em}#1}%
    \else
      \addcontentsline{toc}{part}{#1}%
    \fi
    {\parindent \z@ \raggedright
     \interlinepenalty \@M
     \normalfont\centering
     \ifnum \c@secnumdepth >\m@ne
       \LARGE\bfseries \partname\nobreakspace\thepart
       \par\nobreak
     \fi
     \huge \bfseries #2%
     \markboth{}{}\par}%
    \nobreak
    \vskip 3ex
    \@afterheading}
\renewcommand\partname{Topic}
\makeatother

\newcommand*{\eeps}{{\boldsymbol \epsilon}} % interaction energy vector
\newcommand*{\tK}{{\tilde{K}}}  % interaction matrix

\newcommand*{\ff}{{\mathbf f}} % fraction vector

\begin{document}

\title{Controlling cargo trafficking in multicomponent membranes}
\author{Tine Curk$^{1,2}$, Peter Wirnsberger$^{3}$, Jure Dobnikar$^{1,3}$, Daan Frenkel$^{3}$, An\dj ela~\v{S}ari\'c$^{4,*}$}
\affiliation{$^1$Institute of Physics, Chinese Academy of Sciences, \\ Beijing, China \\$^2$Department of Chemistry, University of Maribor,\\ Maribor, Slovenia\\ $^3$Department of Chemistry, University of Cambridge,\\ Cambridge, UK\\$^4$Department of Physics and Astronomy, Institute for the Physics of Living Systems\\ University College London, London, UK\\}
\email{a.saric@ucl.ac.uk}

%%%%%%%%%%%% BEGIN INTRODUCTION %%%%%%%%%%%
\begin{abstract}
Biological membranes typically contain a large number of different components dispersed in small concentrations in the main membrane phase, including proteins, sugars, and lipids of varying geometrical properties. Most of these components do not bind the cargo. Here, we show that such `inert' components can be crucial for precise control of cross-membrane trafficking. 
Using a statistical mechanics model and molecular dynamics simulations, we demonstrate that the presence of inert membrane components of small isotropic curvatures dramatically influences cargo endocytosis,  even if the total spontaneous curvature of such a membrane remains unchanged. Curved lipids, such as cholesterol, as well as asymmetrically included proteins and tethered sugars can hence all be actively participating in controlling membrane trafficking of nanoscopic cargo. We find that even a low-level expression of curved inert membrane components can determine the membrane selectivity towards the cargo size, and can be used to selectively target membranes of certain compositions. 
Our results suggest a robust and general way to control cargo trafficking by adjusting the membrane composition without needing to alter the concentration of receptors nor the average membrane curvature.  This study indicates that cells can prepare for any trafficking event by incorporating curved inert components in either of the membrane leaflets.
\end{abstract}
\maketitle
\bibliographystyle{apsrev4-1}

\section*{Introduction}
Trafficking of nanoscopic cargo such as viruses and nanoparticles across biological membranes is of central interest for a wide range of phenomena, from pathogen infection, design of synthetic drug-delivery vehicles and imaging agents, to the study of nanoparticle toxicity. Trafficking of cargo that are larger than the thickness of the cell membrane typically involves tight wrapping of the object by the membrane~\cite{bahrami2014wrapping}, followed by scission and budding off. The wrapping can be spontaneous, without any assisting factors, or supported by curvature-inducing endocytotic machinery, including BAR proteins, clathrin, and COPII~\cite{simunovic2015physics,johannes2015building,johannes2014bending,simunovic2016physical}. 

Cellular trafficking necessarily involves crossing of physical barriers, and the physical mechanisms of nanoscopic cargo uptake have been thoroughly studied~\cite{zhang2015physical,kozlov2016membrane}. Much of this research has focused on understanding the physics of nanoparticle wrapping by homogeneous membranes~\cite{lipowsky1998vesicles,gao2005mechanics,zhang2009size,vacha2011receptor,vsaric2012mechanism,huang2013role,dasgupta2013wrapping,dasgupta2014shape,schubertova2015influence,van2016lipid}, such as vesicles of different sizes and shapes, and highly curved membrane segments~\cite{agudo2015critical,agudo2015adhesive,bahrami2016role,zhao2017nanoscale}. 
However, biological membranes are rarely locally homogeneous, and usually contain a large number of different components, which are present in small amounts in the main membrane phase, including lipids of different geometrical properties, embedded proteins, and anchored sugars~\cite{singer1972fluid}. Some of these membrane components can bind specifically to the complementary ligands on the cargo, and are referred to as receptors. Most of the membrane components, however, do not bind to the cargo, and have been thus far considered purely as spectators in the trafficking processes.

\begin{figure*}[ht!]
\centering
\includegraphics[width=0.75\textwidth]{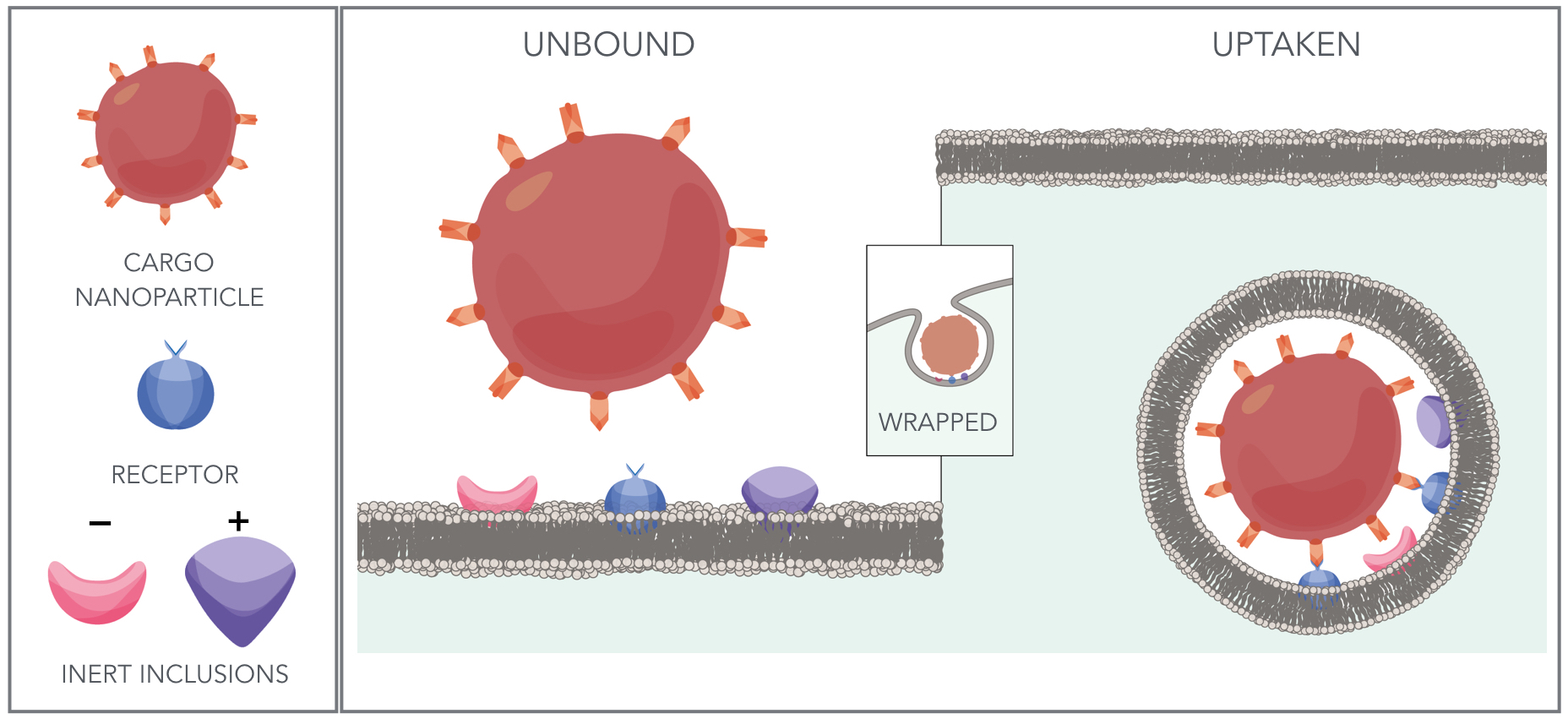} 
\caption{Schematic representation of the cargo-membrane system considered in this paper. The cargo, represented as a spherical nanoparticle, is uniformly covered with ligands, which specifically bind only to receptors in the membrane (coloured in blue). In addition, inert inclusions of isotropic spontaneous positive (purple) or negative curvature (pink) can be expressed in the outer layer of the membrane at varying concentrations. Positive spontaneous curvature locally suppresses the wrapping of particles, while components with negative spontaneous curvature promote the wrapping. The cargo gets tightly wrapped by the membrane upon binding to it, and can be uptaken in a form of a bud.}
\label{fig-scheme}
\end{figure*}

Here we show that non-cargo-binding membrane components of varying geometries can be crucial in controlling cargo trafficking. Using analytical and numerical modelling, we study the effect of small spontaneous curvatures of membrane receptors and inert lipid inclusions on the membrane uptake of nanoscopic cargo, such as nanoparticles, viruses, and other nano-objects (Fig.~\ref{fig-scheme}). Components of small spontaneous curvatures, which are of the same order of magnitude as the curvatures of naturally occurring lipids and inclusions, can be expressed in the way that does not change the total curvature of the membrane, or cause phase-separation and formation of highly curved regions, and still dramatically influence cargo uptake. We present phase diagrams showing how spontaneous curvatures of receptors and inert inclusions affect the onset of endocytosis. Importantly, we show that curved inert membrane inclusions can be used to precisely control cargo trafficking, to selectively target membranes of certain compositions, and to selectively target cargo of certain size, even if the total spontaneous curvature of the membrane remains unchanged. These findings present a robust and general way to control cargo trafficking without altering the ligand-receptor binding properties, and demonstrate that membrane trafficking depends not only on the presence of specific receptors, but also on the overall membrane composition. 

\section*{Theory of cargo uptake in multicomponent membranes}
We start with a theoretical analysis of the free energy cost of crossing a multicomponent membrane via passive endocytosis. We will compute free energies of two limiting cases: when the cargo is not yet in contact with the membrane, and after endocytosis when it is fully wrapped by the membrane and detached from the parent bilayer, as shown in Fig.~\ref{fig-scheme}. 
Let us consider a membrane that contains an arbitrary number of components, which can represent lipids, cargo-binding receptors, or non-interacting membrane inclusions (Fig.~\ref{fig-scheme}). The membrane composition is completely described by a unit vector ${\bm f}=\{f_1,f_2, ... \}$ specifying the density fraction $f_j$ of all distinct components $j$ present in the membrane. Every membrane component $j$ is assumed to have cylindrical symmetry and its size is of the order of a transmembrane receptor and a few surrounding lipids, such that the lateral membrane area per component is $a^2 \approx 25$nm$^2$. Each individual component type is characterised by a spontaneous curvature $c_{0,j}$ and a favourable binding energy to the cargo $-\epsilon_j$ associated with it. $c_{0,j}$ is a partial molar-like quantity defined by a membrane consisting of a pure component $j$.  The sign of the curvature is defined in Fig. 1: if an inclusion is curved towards the cargo it is considered positively curved, while if it is curved away from the cargo its curvature is defined as negative. We assume that bending modulus $\kappa$ and Gaussian stiffness $\bar{\kappa}$ are the same for all components.

The total free energy change upon cargo endocytosis, $\Delta F$, can then be written in terms of individual contributions due to the membrane curvature, $\Delta F_c$, binding to the cargo, $\Delta \epsilon$, the mixing entropy, $\Delta S$, and the lateral membrane pressure $\Pi$:
\begin{equation}
\Delta F  =  \Delta F_c + \Delta \epsilon  - T \Delta S - \Pi A_w  \;,
\label{eq-first}
\end{equation}
where $T$ is the absolute temperature, ans $A_w$ is the membrane area wrapped around the cargo.
Using the mean-field model presented in the Methods section we obtain a closed form expression for the endocytosis free energy:
\begin{equation}
\begin{split}
\Delta F&  = N_{\rm w} \sum_j f_j K_j \bigg[ \epsilon_j + \frac{2\kappa}{R_{\rm w} \rho}\left(\frac{1}{R_{\rm w}} + c_{0,j} \right)  \\
& + k_{\rm B}T \ln(K_j)\bigg]  - \Pi A_{\rm w}  + 4 \pi \bar{\kappa} + N_{\rm w} \mathcal{O}\left(\frac{A_{\rm w}}{A} \right)\;,
\end{split}
\label{eq-delF}
\end{equation}
which is a function of the membrane composition $\bm{f}$, spontaneous curvature vector $\bm{c_0}$, interaction vector $\bm{\epsilon}$, and the radius of the membrane envelope wrapped around the cargo $R_{\rm w}$, $A_{\rm w} = 4\pi R_{\rm w}^2$.
The first term on the right-hand side captures the binding of membrane components to the cargo, the second term is the curvature penalty due to the mismatch of spontaneous curvatures, and the third term captures the effect of the membrane composition change between the flat membrane and the wrapped part with the equilibrium constant
\begin{equation}
K_{j} = \frac{ e^{-\beta \left[ \epsilon_j+ \frac{2\kappa}{R_{\rm w}\rho}(1/R_{\rm w} + c_{0,j}) \right] }} {\sum_{j} f_j \, e^{-\beta \left[ \epsilon_j+ \frac{2\kappa}{R_{\rm w}\rho}(1/R_{\rm w} + c_{0,j}) \right] }} \;.
\label{eq-Kj}
\end{equation}
The prefactor $N_{\rm w} = \rho A_{\rm w} = 4 \pi R_{\rm w}^2 \rho$ is the number of membrane components in the wrapped membrane, with $\rho=1/a^2$ being the overall component number density.  The fourth and fifth terms are the membrane lateral pressure and the Gaussian bending rigidity contribution. Finally,  $\mathcal{O}(A_{\rm w} / A )$ captures all terms which can be neglected in the dilute limit.
\footnote{Note that the free energy difference~\eqref{eq-first} is related to the equilibrium ratio of the cargo densities on the two sides of the membrane: $\Delta F/k_{\rm B}T = -\ln \left( \frac{\rho_{\rm cargo,in}}{\rho_{\rm cargo,out}}\right)$. For high cargo concentrations one should use the ratio of fugacities.}

\subsection*{Analytical results}

\begin{figure}[ht!]
\centering
\includegraphics[width=1\textwidth]{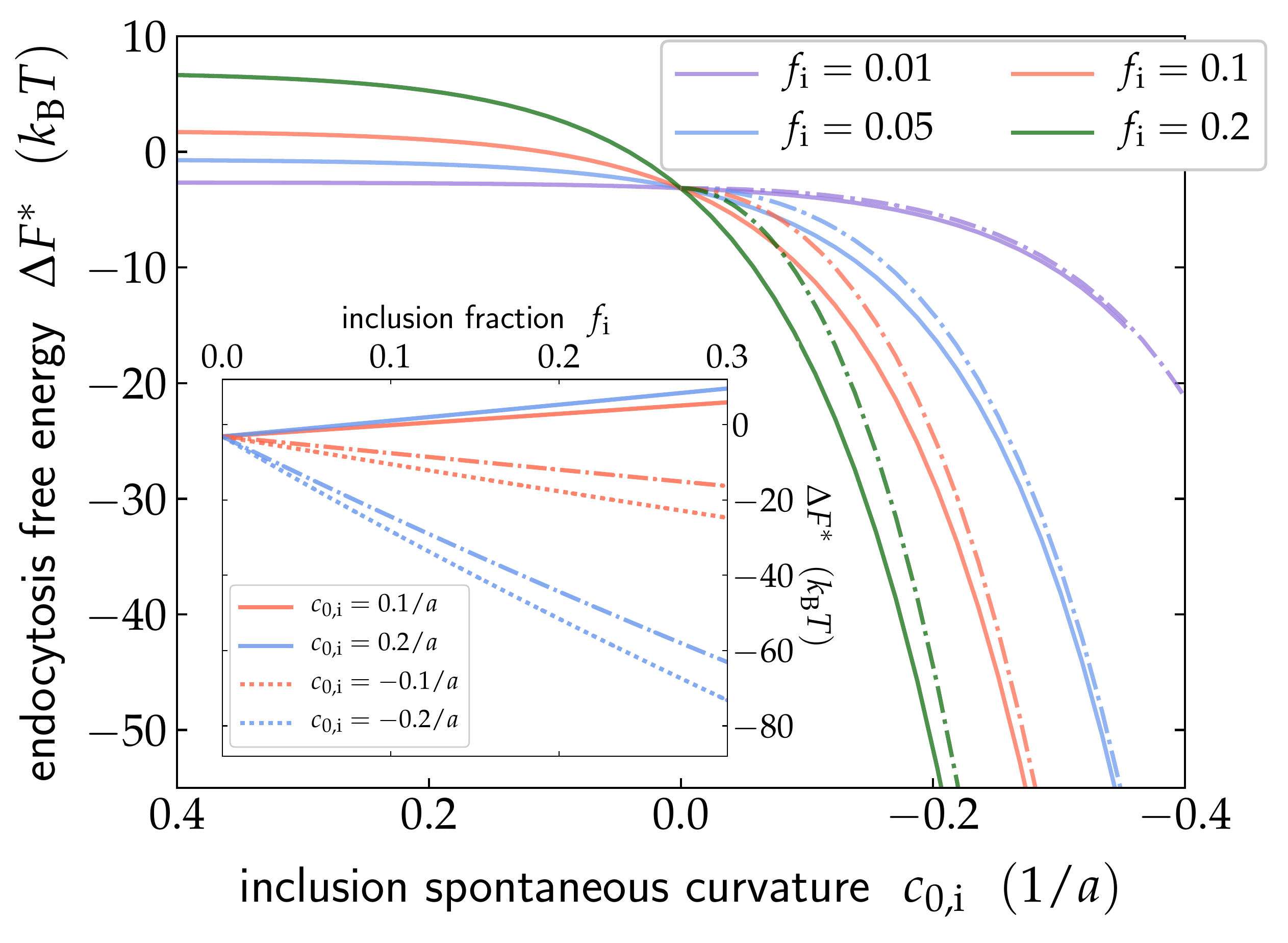}
\caption{ \textbf{The effect of inert inclusions on the endocytosis free energy.}  Changing the spontaneous curvature $c_{\rm 0,i}$ of inert inclusions dramatically influences the free energy cost of crossing the membrane. Inset: Varying the fraction of inclusions  $f_{\rm i}$  at constant spontaneous curvature. The dot-dashed curves on both plots show the case of two types of inclusions with opposite spontaneous curvatures $c_{\rm 0,i}=-c_{\rm 0,i'}$ and $f_{\rm i'}=f_{\rm i}$, such that the total membrane spontaneous curvature is zero. Parameters: $R_{\rm w} = 5a$, $f_{\rm r}=0.1$, $\epsilon_{\rm r} = -4k_{\rm B}T$, $\kappa = 23 k_{\rm B}T$, where $a$ is the length-scale corresponding to the membrane thickness, $a \approx 5$nm.  Note that $\Delta F^* = \Delta F - 4 \pi \bar{\kappa} + \Pi A_{\rm w}$.}
\label{fig-th}
\end{figure}

In what follows we focus on the effects of membrane composition on the passive uptake of cargo. Figure~\ref{fig-th} shows how the endocytosis free energy changes when inert components of varying spontaneous curvatures are included into the membrane. The free energy can be shifted by a few tens of $k_{\rm B}T$ even when the fraction of inert inclusions is only a few percent. Inclusions of negative spontaneous curvatures (see Fig.~\ref{fig-scheme} for the definition) lower the free energy change for cargo uptake, while the positive spontaneous curvature displays the opposite effect. The effect of promoting endocytosis by inclusions of negative spontaneous curvature is much stronger than the corresponding suppressing effect of inclusions with positive spontaneous curvature. This asymmetry is caused by the recruitment of inclusions of desirable curvature to the wrapped part of the membrane, an effect that is similar to the recruitment of receptors that bind to the cargo. 

The addition of non-interacting inclusions of spontaneous curvature can substantially shift the free energy of endocytosis, and can be used as a mechanism to control the cargo uptake. Crucially, the effect remains even when both positive and negative inclusions are present such that the total spontaneous curvature of the multicomponent membrane is kept at zero, as displayed by dot-dashed lines in Fig. ~\ref{fig-th}.

\section*{Computer simulations of cargo uptake in multicomponent membranes}
To test the predictive quality of our theoretical considerations, we turn to computer simulations and explore in depth the effect of membrane composition and geometry of its components on cargo uptake. Computer simulations fully capture the possible existence of stable,  partially wrapped states~\cite{zhang2009size,agudo2015critical,agudo2015adhesive}, which are ignored in our simple analytical model, and also allow for the analysis of the local membrane composition during the endocytosis, giving a deeper insight into the process.

The membrane is modelled using a coarse-grained one-particle thick model~\cite{yuan2010one}. As shown in Fig.~\ref{fig-sim}, the membrane is composed of `lipid' beads of zero spontaneous curvature, the cargo-binding beads which we call receptors, and inert (non-cargo binding) beads that model membrane inclusions and can carry a spontaneous curvature. %All the membrane components are of the same size and mass; the membrane lipid beads are denoted with the index 'm', the receptor beads with 'r', and the curvature-inducing inert inclusions are denoted with 'i'. 

\subsection{Cargo endocytosis at varying membrane compositions}

\begin{figure*}[ht!]
\center
\subfigure[]{\includegraphics[width=0.46\textwidth]{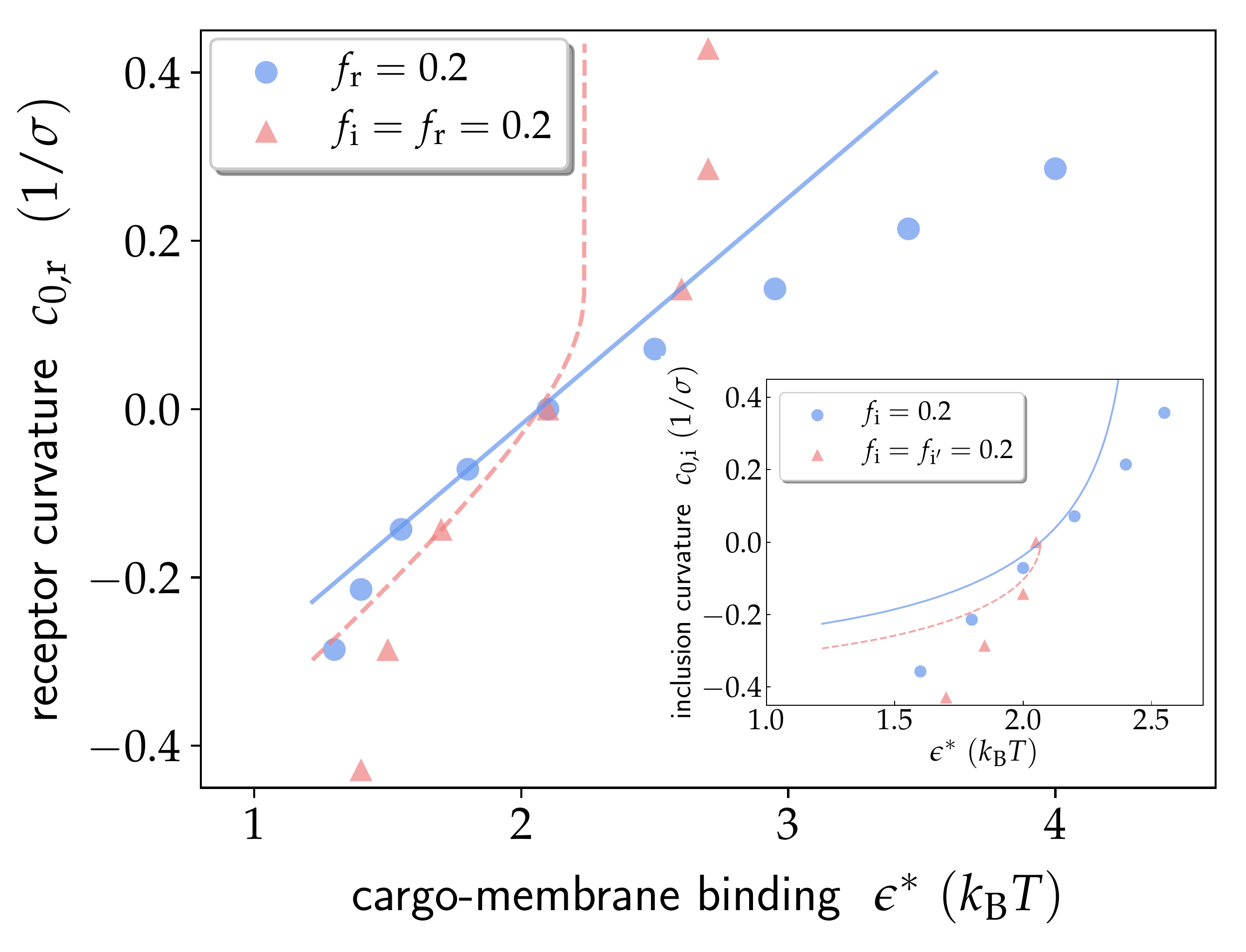}} \ \ \
\subfigure[]{\includegraphics[width=0.45\textwidth]{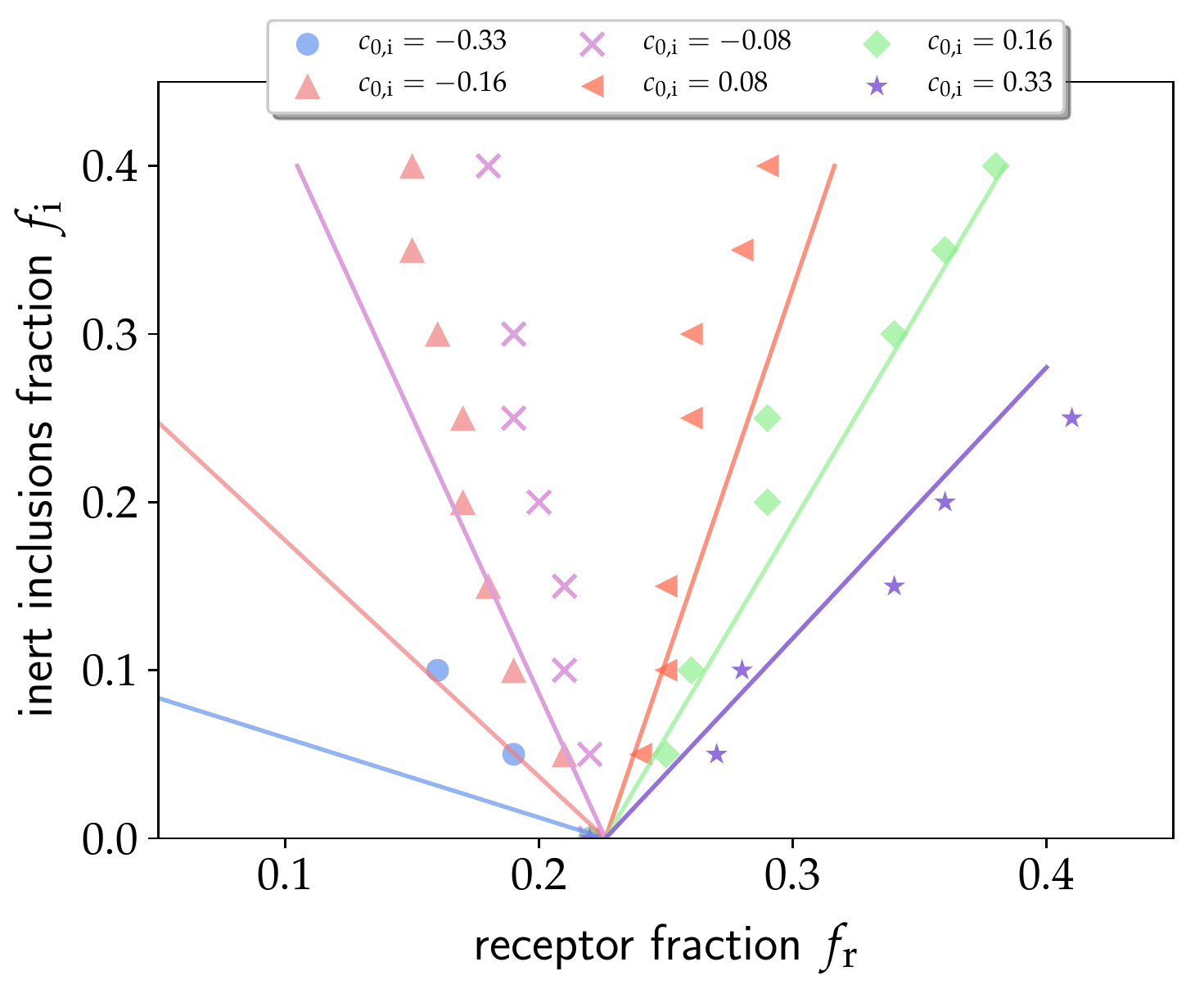}}
\caption{(a) \textbf{Spontaneous curvatures of receptors and inert inclusions control endocytosis.} Symbols denote simulation results and lines (solid and dashed) correspond to predictions of the mean-field theory. Phase diagram showing the dependence of the onset of endocytosis on the receptor curvature, at constant receptor and inclusion fraction. Inset: Dependence of the onset of endocytosis on the curvature of inert inclusions at $c_{\rm 0,r}=0$. Pink dashed lines and pink triangles show the case where the total spontaneous curvature of the membrane remains zero: $c_{\rm 0,i} = -c_{\rm 0,r}$ in the main plot, and two inclusion types $c_{\rm 0,i'} = -c_{\rm 0,i}$ in the inset. (b) \textbf{Tuning the onset of endocytosis by the membrane composition.} Phase diagram showing how the fraction $f_{\rm r} $ of receptor beads needed for the onset of endocytosis changes with the addition of inert beads of fraction $f_{\rm i}$ and spontaneous curvature $c_{0,{\rm i}}$. Receptors have zero spontaneous curvature $c_{0,{\rm r}}=0$ and $\epsilon^*=2k_{\rm B}T$. Solid lines show the theoretical prediction at $\Delta F=0$. Parameters used in the analytical calculations in (a) and (b): $R_{\rm p}=8 \sigma$, $R_{\rm w} = R_{\rm p} + a$, $a=\sigma/\sqrt{1.21}$, $\kappa = -\bar{\kappa} = 22 k_{\rm B}T$, $\Pi = 0$, $\epsilon_{\rm r} = -\epsilon^*+1.24k_{\rm B}T$.}
\label{fig-sim1}
\end{figure*}

We first consider receptors of varying spontaneous curvatures, and measure the cargo-receptor interaction energy at the onset of the cargo endocytosis, $\epsilon^*$, as a function of the receptor curvature. Fig. \ref{fig-sim1}(a) predicts that this onset of endocytosis can be dramatically shifted if the receptors posses non-zero spontaneous curvature. The onset can be decreased by over 2$k_{\rm B}T$ per receptor when going from a receptor of positive spontaneous curvature ($c_{0,\rm r}=-0.16\sigma^{-1}\approx -0.03 {\rm nm}^{-1}$) to a receptor of a negative spontaneous curvature ($c_{0,\rm r}=0.16\sigma^{-1}\approx 0.03 {\rm nm}^{-1}$)~\footnote{Assuming membrane bead size of $\sigma = 5$nm.}, Fig.~\ref{fig-sim1}(a). This result is in line with the previously reported results of analytical calculations for nanoparticle uptake by homogeneously adhesive vesicles of varying bilayer asymmetries~\cite{agudo2015critical}. 

More interestingly, when inert membrane inclusions are added (Fig.~\ref{fig-sim1}(a) and the corresponding inset), be it lipids or proteins, at a constant receptor concentration, the onset of endocytosis is also significantly altered. The effect remains when both positive and negative components are present such that the total spontaneous curvature of the multicomponent membrane is kept at zero, shown by the pink triangular symbols in Fig.~\ref{fig-sim1}(a) and its inset. In this case the cargo simply `recruits' the desirable components from the membrane `reservoir'.

Fig.~\ref{fig-sim1}(b) presents a comprehensive phase diagram depicting how the addition of inert inclusions of varying spontaneous curvatures shifts the onset concentration of receptors needed for endocytosis.
The ability to tune the cargo trafficking by incorporating membrane inclusions can be used as a strategy in trafficking of cargo; the level of expression of generic inclusions of a non-zero spontaneous curvature can be a way to enhance, or prevent, the endocytosis and exocytosis of nanoparticles and pathogens. %As shown in Fig.~\ref{fig-sim1}(b), an over-expression of curved inert inclusions can significantly change the receptor concentration needed for cargo uptake, from over 40\% to only 10\%.

We will now compare the results from simulations and our analytical theory. To obtain a theoretical phase diagram using the analytical theory we assume $\Delta F=0$ in (Eq.~\ref{eq-first}) ~\footnote{In principle, the onset $\Delta F = -k_{\rm B} T \ln \left( \frac{\rho_{\rm cargo,in}}{\rho_{\rm cargo,out}}\right)$ is determined by the equilibrium ratio of cargo concentrations on the two sides of the membrane}, and numerically compute the curvature ${\mathbf c}_{0,j} ( \epsilon_{\rm r} | \Delta F= 0)$ for a given value of receptor-cargo binding energy $\epsilon_{\rm r}$ (Fig.~\ref{fig-sim1}(a)), or inert inclusion fraction ${\mathbf f}_{\rm i} ( f_{\rm r} | \Delta F= 0)$ for a given value of receptor fraction $f_{\rm r}$ (Fig.~\ref{fig-sim1}(b)), while keeping all the other parameters constant. \footnote{Note that $\epsilon_{\rm r} \ge 0$ should be negative, otherwise vesicle formation is more thermodynamically stable than cargo wrapping, see SI for the discussion on membrane stability and spontaneous vesicle formation.} The lines in Fig.~\ref{fig-sim1}(a) and (b) depict the results of the analytical theory. The analytical theory results show the same qualitative trends as the results found in simulations, although deviations can also clearly be seen. This discrepancy between simulations (symbols) and theory (lines) in Fig.~\ref{fig-sim1} is mainly due to the stability of partially bound states that are not considered by the analytical theory, and the corresponding free energy barrier for the complete cargo wrapping, which can in turn delay endocytosis~\footnote{In the mapping between simulation and analytical results, the receptor interaction is a sole fitting parameter because the theory implicitly assumes a square well-like interaction while the simulation model uses a Lennard-Jones potential. We found that $\epsilon_{\rm r, theory} = -\epsilon^*_{simulations} + 1.24k_{\rm B}T$.}. 

\subsection{Super-selectivity to membrane composition}
Biological membranes need to be highly selective when allowing for cargo trafficking, to ensure robust functioning of the cell. 
A hallmark of such a super-selective targeting is a sharp increase in the cargo binding upon a small change in the membrane composition. Such a behaviour results in a very low efficiency of the uptake below a  threshold composition, while around the threshold the uptake sharply increases. Given that the endocytosis uptake sensitively depends on the presence of curved inclusions, it is tempting to assume that expressing such inclusions (e.g. cholesterol molecules) is one of nature's ways to tune the selectivity of  membranes to specific cargo.

\begin{figure*}[ht!]
\centering
\subfigure[]{\includegraphics[width=0.46\textwidth]{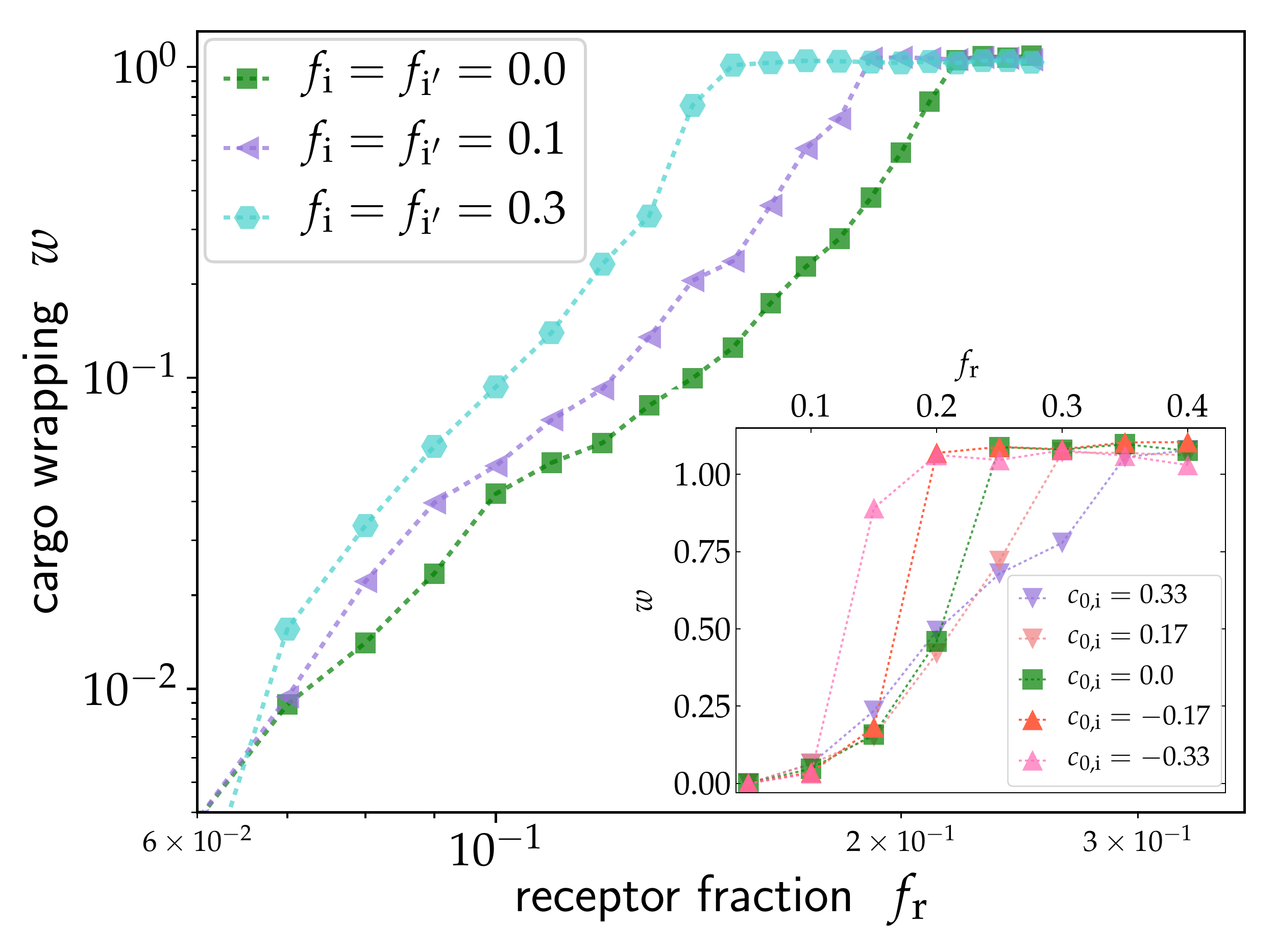}} \ \ \
\subfigure[]{\includegraphics[width=0.45\textwidth]{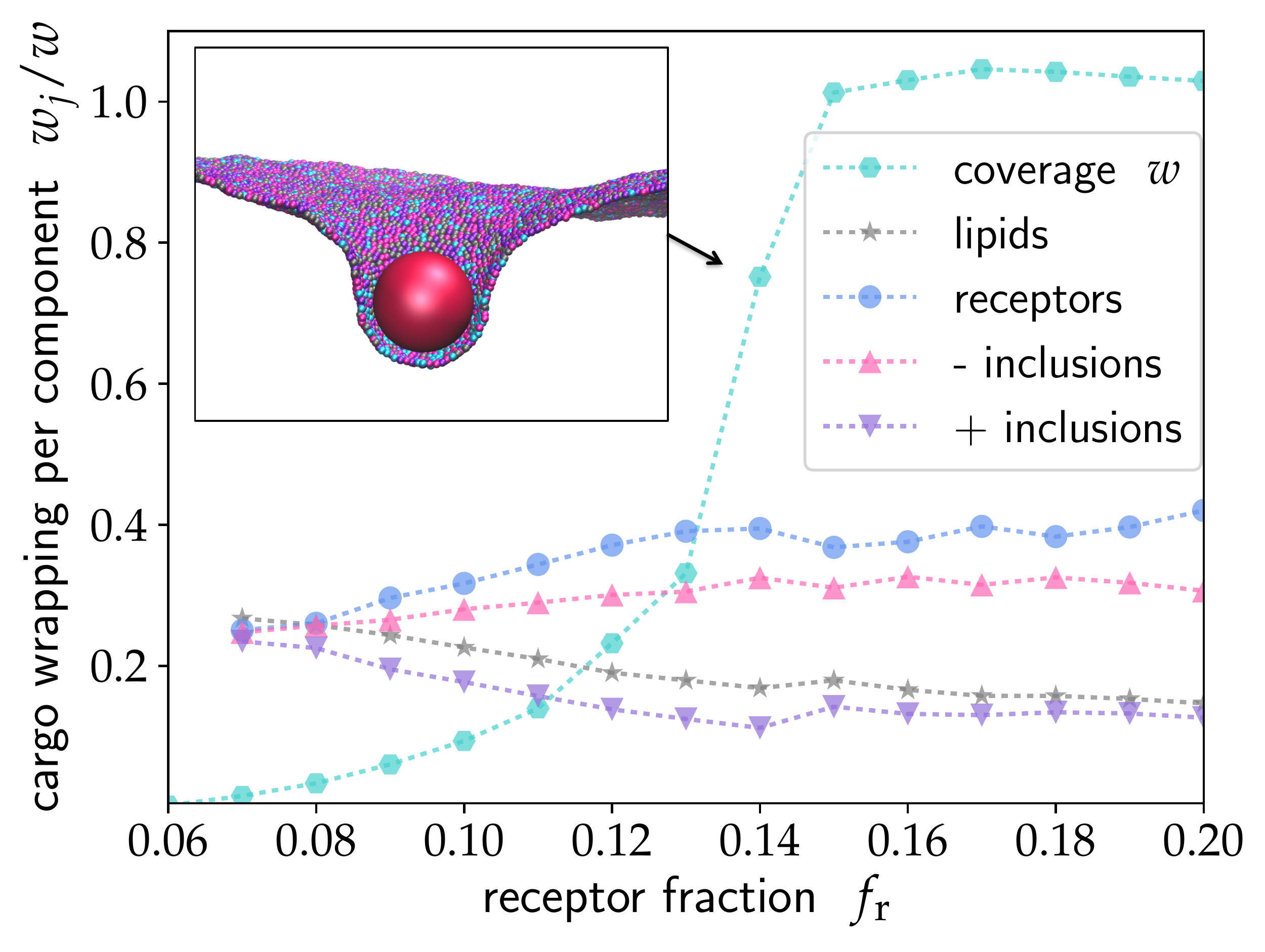}}
\caption{\textbf{Spontaneous curvatures of inert membrane components control the selectivity of cargo towards membranes of different compositions.} (a) Dependence of the cargo wrapping by the membrane (Eq.(~\ref{eq-fwrap})) on the fraction of receptors upon addition of inert beads only of varying spontaneous curvatures ($f_{\rm i}=0.2$, inset), or a mixture ($f_{\rm i} = f_{\rm i'}$)  of inert beads of both positive and negative spontaneous curvatures  $c_{0,{\rm i}} = - c_{0,{\rm i'}} = 0.33/\sigma$  (main figure). The coverage $w$ above 1 indicates full endocytosis.  (b) The fractional coverage of the cargo particle by different membrane component types, $w_j/ w$, at $f_{\rm i} = f_{\rm i'}=0.3$. The cargo spontaneously recruits receptors and negatively curved inclusions, while excluding membrane 'lipid' components and positively curved inclusions.  The receptor fraction is $f_{\rm r}=0.2$, the cargo nanoparticle radius is $R_{\rm p}=8 \sigma$ and $\epsilon^*=2 k_{\rm B}T$. Inset shows the configuration snapshot at $f_{\rm r}=0.14$, the color scheme corresponds to symbol colors used in Fig.~\ref{fig-scheme}. 
}
\label{fig-SS}
\end{figure*}

Here we explore the effect of inert inclusions on the selectivity of cargo uptake. We find that the sensitivity of the membrane to the cargo uptake increases with the addition of negatively curved inclusions, but decreases for positively curved inclusions (inset in Fig.~\ref{fig-SS}(a)). The sharpness of the transition is governed by the stability of partially wrapped states, which is enhanced when positively curved inclusions are present, in line with a previous analytical study~\cite{agudo2015adhesive}. Moreover, Figure~\ref{fig-SS}(a) shows that the selectivity increases when inclusions with both positive and negative spontaneous curvature are added, such that the total membrane spontaneous curvature remains zero. This asymmetry arises due to recruiting of the inclusions with negative spontaneous curvature to the cargo, while inclusions with positive spontaneous curvature are expelled from the wrapped membrane area, as shown in Fig. ~\ref{fig-SS}(b). The effect is the same for both endocytosis and exocytosis.

A key message of these results is that for cellular trafficking it is better to create a membrane with a constant (i.e zero) curvature from a mixture of components with opposite signs, than from homogeneous components of uniform zero curvature. Such a heterogeneous membrane is then prepared for a plethora of trafficking effects. Thus  our findings suggest a generic  functional role of the ubiquitous non-cargo-binding curved lipids ~\cite{holopainen2000vectorial, mcmahon2005membrane,roux2005role,martens2008mechanisms,kamal2009measurement,sorre2009curvature}.

\section*{Kinetics of cargo uptake: selectivity to cargo size}
In experiments, as well as in molecular dynamics simulations, what is typically probed is the amount of endocytosis in a finite time. Hence, the kinetics, and not only thermodynamics, of the cargo engulfment matters. Many experimental studies have reported the existence of an optimal size of nanoparticles for which the rate of endocytosis is the highest, while it becomes slower for lower and larger nanoparticle sizes~\cite{chithrani2006determining,jiang2008nanoparticle}. Several theoretical studies have rationalised this non-monotonic behaviour by a competition between thermodynamics, which disfavours uptake of small nanoparticles, and diffusion-limited recruitment of receptors to the cargo~\cite{gao2005mechanics}, which disfavours large nanoparticles. Similar results can be recovered by considering the competition between thermodynamic driving forces related to the creation of the bud neck and frictional forces of cargo wrapping~\cite{agudo2015critical}. When many large nanoparticles are considered, depletion of receptors can also occur~\cite{zhang2009size}.

\begin{figure}
\centering
\includegraphics[width=0.95\textwidth]{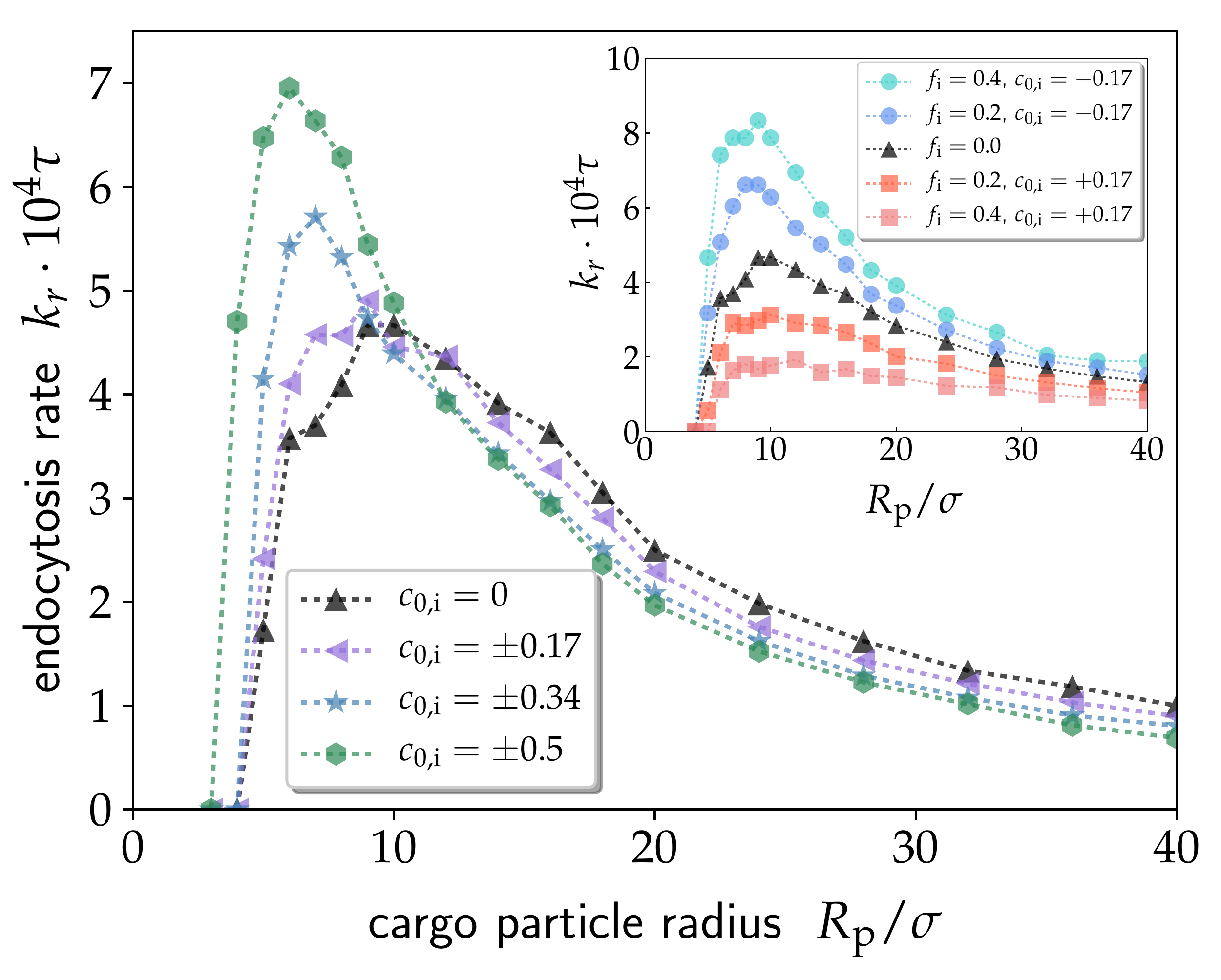} \\
\caption{\textbf{Endocytosis rate $k_{\rm r}$ depends non-monotonically on the cargo size.}  Presence of inert inclusions of both positive and negative spontaneous curvatures sharpens the membrane selectivity towards the cargo size.  Inset: Addition of components with just negative (positive) spontaneous curvature increases (decreases) the rate. Fraction of receptor beads is fixed at $f_{\rm r}=0.4$, interaction $\epsilon^*=2.5k_{\rm B}T$ and curvature $c_{0,{\rm r}}= 0$.  In the main plot the fraction of inclusions is $f_{\rm i} = f_{\rm i'}=0.2$ with opposite spontaneous curvatures $c_{0,{\rm i}}= - c_{0,{\rm i'}}$. The scaling in the limit of large cargoes follows an inverse power law $k_{\rm r} \sim 1/R_p^{\gamma}$ with the exponent $\gamma \approx 1.5$ indicating an intermediate regime between friction ($\gamma=1$) and diffusion ($\gamma=2$) limited endocytosis, see SI.}
\label{fig-chR}
\end{figure}

Here, we examine this process by explicitly measuring the rate of engulfment of nanoparticles of various sizes into multicomponent membranes. As shown in Fig. \ref{fig-chR}, we  recover the non-monotonic dependence of the rate of uptake on the cargo radius. Moreover, the value of the rate as well as the selectivity of the membrane uptake towards cargo of certain radii, can be altered by the presence of curved inert inclusions (Figure~\ref{fig-chR}). Interestingly, the presence of both positively and negatively curved inert inclusions sharpens the selectivity towards the cargo size, shifting it towards the lower values. On the other hand, presence of only positively/negatively curved inclusions decreases/increases the endocytosis rate for all cargo sizes as shown in the inset of Fig.~\ref{fig-chR}. 

This observation can be rationalised in the following way: for large cargoes the mean curvature in the neck is positive, therefore, inclusions with positive spontaneous curvature occupy and stabilise the neck region, as shown in Fig.~\ref{fig-neck}. This region then  presents a barrier for receptors and negatively curved inclusions to diffuse to the membrane reducing the endocytosis rate. Conversely, for small cargoes the mean curvature in the neck vanishes, and the curved inclusions do not occupy the neck region. In this case, negatively curved inclusions can freely diffuse to the membrane area that wraps the cargo, decreasing the free energy for endocytosis, and enhancing its rate (see Supplementary Information). 

Our analysis shows that the presence of inert inclusions can be used not only for selecting cargo of certain membrane binding properties, but also for selecting cargo of specific sizes. This property of multicomponent membranes can be readily utilised in the design of nano-vehicles for targeted delivery of chemicals.

%\begin{figure}
%\centering
%\includegraphics[width=0.95\textwidth]{Fig6+inset_feb18} \\
%\caption{\textbf{Endocytosis rate $k_{\rm r}$ depends non-monotonically on the particle size.}  Presence of inert inclusions of both positive and negative spontaneous curvatures sharpens the membrane selectivity towards the cargo size. Inset: Addition of components with negative (positive) spontaneous curvature increases (decreases) the rate. Fraction of receptor beads is fixed at $f_{\rm r}=0.4$, interaction $\epsilon^*_{\rm r}=2.5k_{\rm B}T$ and curvature $c_{0,{\rm r}}= 0$.  In the main plot the inclusion fraction is $f_{\rm i} = f_{\rm i'}=0.2$ with opposite spontaneous curvatures $c_{0,{\rm i}}= - c_{0,{\rm i'}}= 0$. To capture the correct dynamics, the mass and friction coefficient of the membrane beads is set to unity, while the cargo nanoparticle parameters are rescaled accordingly: mass of nanoparticle is $m_{\rm p} = 8 (R_{\rm p} / \sigma)^3$ and nanoparticle friction coefficient $\gamma_{\rm p}=2 R_{\rm p} / \sigma $. The rate $k_{\rm r}$ is expressed in units of 1/$\tau$. The scaling in the limit of large particles follows an inverse power law $k_{\rm r} \sim 1/R_p^{\gamma}$. For $c_{0,{\rm i}}= 0$ the exponent is $\gamma \approx 1$ indicating that friction, and not receptor diffusion, is the reason for the rate slowing down. However, the scaling exponent increases to $\gamma \approx 1.4$ at $c_{0,{\rm i}} = - c_{0,{\rm i'}} =0.5/\sigma$ indicating an increased role of component diffusion, see SI.}
%\label{fig-chR}
%\end{figure}

\section*{Discussion and Conclusions}
By considering interactions of nanoscopic cargo with multicomponent membranes, we have shown that the necessary conditions for the cargo uptake, namely the critical concentration of the membrane receptors, and the critical binding energy between the membrane receptors and the cargo ligands, can be precisely tuned by the overall membrane composition. In particular, for a given ligand-receptor pair, including only 10\% of inert components of  spontaneous negative curvatures, as small as 0.03nm$^{-1}$, can decrease the concentration of receptors needed for endocytosis by $\sim 15$\%, enabling easier cargo uptake. Such small local spontaneous curvatures fall into the range of curvatures of typical membrane components; for instance the experimentally determined spontaneous curvature of DOPC in DOPE is $-0.05$ nm$^{-1}$, while that of cholesterol is $-0.3$ nm$^{-1}$~\cite{martens2008mechanisms}. Conversely, asymmetrically expressing inert inclusions of positive spontaneous curvatures decreases endocytotic efficiency and can possibly protect cells from entry of pathogens and other undesirable nano-objects. For comparison, common lyso-phospholipids exhibit positive spontaneous curvatures in the range of $0.02-0.25$ nm$^{-1}$~\cite{kamal2009measurement}. Therefore, the role of curved lipids, such are cholesterol, in controlling membrane physical properties possibly extends beyond adjusting the membranes fluidity and bending rigidity. Furthermore, the presence of inclusions of negative spontaneous curvatures increases  membrane selectivity towards the cargo nature and the cargo size (Fig.~\ref{fig-SS} and Fig.~\ref{fig-chR}), and increases overall specificity of the trafficking processes.
\\
Our results are in a good qualitative agreement with previous analytical calculations that considered the role of the bilayer asymmetry in controlling nanoparticle engulfment for homogeneous adhesive membranes~\cite{agudo2015critical}. Recently, computer simulations have shown that the phase diagram and dynamics of membrane tubulation induced by BAR domains, anisotropic membrane-curving proteins, can be modified by the presence of curved inclusions~\cite{noguchi2017acceleration}, which is also in line with our results. %Although the authors study phase-separation of membrane inclusions themselves, and not cargo uptake, our results are in line with the findings reported in~\cite{noguchi2017acceleration}. 
Importantly, distinct from previous studies, our results hold even if the sum of spontaneous curvatures of membrane components remains unchanged or equals to zero. 

We have previously studied how adsorption of multivalent particles onto rigid surfaces with many receptors of different binding properties can be controlled by the receptor composition~\cite{curk2017optimal}. Here we show that small concentrations of inert membrane inclusions of negative spontaneous curvature can dramatically influence the membrane selectivity towards cargo engulfment, rendering the expression of inert inclusions an attractive general mechanism in controlling cell trafficking. 

Moreover, we demonstrated that the presence of negatively curved inert inclusions increases the sensitivity towards the cargo size, while the positively curved inclusions wash away this effect. Our results suggest that interactions of nano-objects with biological membranes, which are inherently inhomogeneous, display rich behaviour that goes well beyond the usually considered ligand-receptor interactions. We provide a novel and general route for modulating cargo trafficking in biological and synthetic membranes, and selectively targeting membrane composition.

\section*{Acknowledgments}
\noindent We acknowledge fruitful discussion with Giuseppe Battaglia, as well as support from the Herchel Smith scholarship (T.C.), the CAS PIFI fellowship (T.C.), the UCL Institute for the Physics of Living Systems (T.C. and A.\v{S}.), the Austrian Academy of Sciences through a DOC fellowship (P.W.), the European Union's Horizon 2020 programme under ETN grant 674979-NANOTRANS (J.D., D.F.), the Engineering and Physical Sciences Research Council (D.F. and A.\v{S}.), the Academy of Medical Sciences and Wellcome Trust (A.\v{S}.), and the Royal Society (A.\v{S}.). 

\section{Methods}

\subsection{Theoretical model derivation}

For simplicity, we assume the cargo is spherical. The free energy cost for the engulfment of such a nano-object is composed of the free energy of membrane curvature, the binding energy of the cargo to the membrane components, and the mixing entropy of membrane components.  

The curvature free energy density of the membrane is in general given by the Helfrich Hamiltonian~\cite{helfrich1973elastic}
\begin{equation}
H_{bend} = \frac{\kappa}{2} (2C - c_{0})^2 + \bar{\kappa} K \;,
\label{eq-fc}
\end{equation}
with $C$ and $K$ being the local mean and Gaussian curvature of the membrane, respectively, and $c_0$ the spontaneous curvature. 
We generalise the Helfrich Hamiltonian to capture a multicomponent membrane. Within the homogeneously mixed, mean-field, approximation (see SI) the above equation applies to each component individually, hence
\begin{equation}
F_{c,j}/a^2 = \frac{\kappa_{j}}{2} (2C - c_{0,j})^2 + \bar{\kappa_j} K \;,
\label{eq-fi}
\end{equation}
is the free energy per individual component of lateral size $a^2$ and lateral extension $a \approx 5$nm.
The total mean-field curvature free energy $F_c$ of a membrane patch with mean curvature $C$ can thus be obtained by simply summing over all the component types (SI), such that:
\begin{equation}
\begin{split}
F_c &=  a^2 \sum_{j} f_j F_{c,j} \\
&=  a^2 \sum_{j} f_j \left[ \frac{\kappa_j}{2} (2C - c_{0,j})^2 + \bar{\kappa_j} K \right] \,.
\label{eq-Fc}
\end{split}
\end{equation}

Assuming that the membrane is in a fluid state the mixing entropy of components is given by the Gibbs expression for the entropy per individual component
\begin{equation}
s = - k_{\text B} \sum_j f_j \ln(f_j) \;,
\label{eq-ent}
\end{equation} 
with $k_{\text B} $ the Boltzmann constant.

Finally, in addition to the membrane deformation and mixing entropy of membrane receptors, each membrane component  interacts with the cargo with an energy $-\epsilon_j$. We set this value to zero for all membrane components but the receptors and assume that the cargo is uniformly covered with ligands complementary to the receptors. 

Initially, the membrane is flat with a total area $A$ and composition $\bm{f}$. Upon endocytosis a portion of the membrane area $A_{\rm w}=4\pi R_{\rm w}^2$ is wrapped around the cargo with membrane shell radius $R_{\rm w}$, while the rest $\tilde{A}=A-A_{\rm w}$ is assumed to remain flat. Due to our definition of spontaneous curvature (Fig.~\ref{fig-scheme}) the curvature of a shell is $C=-1/R_{\rm w}^2$ for endocytosis, and $C=1/R_{\rm w}^2$ for exocytosis. The composition of the wrapped part $ \bm{f}^{\rm w}$ is in general different from the remaining flat membrane $\tilde{\bm{f}}$. 
This leads to two conservation laws: the area conservation $A = \tilde{A} + A_{\rm w}$,  and the conservation of membrane material (number of membrane components) $\bm{f} A = \tilde{ \bm{f}} \tilde{A} +  \bm{f}^{\rm w} A_{\rm w}$.

For simplicity we shall assume that all components have the same bending rigidity  $\kappa_j = const. $ and Gaussian bending stiffness $\bar{\kappa}_j = const.$  Furthermore, we assume that the wrapped membrane part is small compared to the total membrane $A_{\rm w} \ll A$ (dilute limit), and that lateral component diffusion is fast compared to the endocytosis timescale. Under these conditions the membrane components can be treated as independently adsorbing to the cargo.  
The dynamical process of endocytosis is rather complicated, however, we are only interested in the free energy change between the initial and the final state. We assume that during all of the endocytosis process individual components are in contact with the remainder of the membrane (the reservoir).  Therefore, the composition of the membrane components that are wrapped around the cargo is given by the generalised Langmuir adsorption isotherm:
\begin{equation}
f^{wrapped}_j = \frac{e^{\beta (\tilde{\mu}_j - F_{c,j}(R_{\rm w}) - \epsilon_j)}}{\sum_{j} e^{\beta (\tilde{\mu}_j - F_{c,j}(R_{\rm w})- \epsilon_j)} } \;.
\end{equation}
$F_{c,j}(R_{\rm w})$ is the curvature free energy of the membrane component that is wrapped around the cargo with radius $R_{\rm w}$, as given by Eq. \eqref{eq-fi}, and $\tilde{\mu}$ is the chemical potential of the membrane component of type $j$ on the flat membrane area $\tilde{A}$.  This chemical potential is defined as:
\begin{equation}
\tilde{\mu}_j = k_{\rm B}T \ln (\tilde{f}_j) + F_{c,j}^{0}\;,
\end{equation}
where $F_{c,j}^0 \equiv F_{c,j}(R_{\rm w} \to \infty)$ denotes the curvature free energy of component $j$ in a flat membrane. 

Using these relations we compute the free energy change upon endocytosis, Eqs. (\ref{eq-first}-\ref{eq-Kj}), detailed procedure is provided in the Supplementary Information. 

\subsection{Simulation Model}

The membrane is modelled using a coarse grained one-particle thick model~\cite{yuan2010one}, which captures membrane fluidity and elastic properties, allows for implementation of the spontaneous curvature of individual membrane components $c_{0,j}$, and permits topological changes, such as budding.  In short, in this model each membrane bead is described by its position and an axial vector. The beads interact with a combination of an attractive potential that depends on the inter-bead distance and drives the membrane self-assembly, and an angular potential that depends on the angle between the axial vectors of neighbouring beads and mimics membrane bending rigidity. The spontaneous curvature per bead is implemented via a preferred angle between two axial vectors, $\theta_0$, and is related to it via $c_0 \approx 2\sin(\theta_0/d_0)$. $d_0\approx r_{\rm cut}^{\rm bead-bead}$ is the average distance between membrane beads, for a tensionless membrane the average distance will be located roughly at the minimum of the attractive potential. Following the notation from the original paper~\cite{yuan2010one}, we choose the parameters $\epsilon_{\rm bead-bead}=4.34k_{\rm B}T, \xi=4, \mu=3, r_{\rm cut}^{\rm bead-bead}=1.12\sigma$ for all the membrane components, which positions our membrane in the fluid phase of the phase space.  Using thermodynamic integration and theoretical considerations we  determined the bending and Gaussian rigidity of this membrane model, $\kappa = -\bar{\kappa} = 22k_{\rm B}T$, see SI for details of the calculation. The mixing curvature terms between membrane beads $i,j$ of different spontaneous curvatures are assumed to be symmetric $c_{0,ij}=\frac{c_{0,ii}+c_{0,jj}}{2}$, and hence, the spontaneous curvature does not lead to phase separation of components of different curvatures.

The cargo nanoparticle interacts with the membrane receptors via a shifted Lennard-Jones potential, where the cargo-receptor interaction strength is controlled by the binding affinity $\epsilon^*$. The interaction between the cargo and any of the non-receptor beads is governed only by volume exclusion described by the Weeks-Chandler-Anderson potential. 
The relevant parameters in our simulations are: the cargo particle radius $R_{\rm p}$,  the spontaneous curvature of receptors beads $c_{0,\rm r}$ and inert inclusions $c_{0,\rm i}$, and the fraction of receptors and inert inclusions in the membrane, $f_{\rm r}$ and $f_{\rm i}$, respectively.  

We simulated a flat square portion of a membrane made of 49920 beads with periodic boundary conditions in a $N\Pi H$ ensemble with lateral pressure $\Pi=-10^{-4}\epsilon/\sigma^3$ to model a nearly tensionless membrane scenario. All the particles in the system are in addition subject to random noise implemented via the Langevin thermostat with friction coefficient set to unity: $\gamma = m /\tau$, where $m$ is the bead mass (set to unity) and $\tau$ the simulation unit of time. To capture the correct dynamics the cargo nanoparticle parameters are rescaled accordingly: mass of nanoparticle is $m_{\rm p} = 8 (R_{\rm p} / \sigma)^3$ and nanoparticle friction coefficient $\gamma_{\rm p}=2 R_{\rm p} / \sigma $.

The initial condition of all simulations is a flat membrane with beads arranged on a hexagonal lattice with a randomly chosen permutation of bead identities (types) and the location of the cargo particle's centre of mass  $R_{\rm p} + 2\sigma$ above the membrane.
All quantities are expressed in terms of the membrane bead diameter $\sigma$, which corresponds to $\sigma=5
$nm in physical units. The bead density for a flat tensionless membrane was measured to be $\rho_{\rm beads} = 1.21\sigma^{-2}$, which provides a mapping with a theoretical component size $a=\sigma/\sqrt{1.21}$. 

The endocytosis is monitored through the wrapping coverage of the cargo by the membrane beads, where the wrapping is defined as:
\begin{equation}
w_j=\frac{N^{\rm contact}_j \sqrt{3}}{8\pi (R_{\rm p}/\sigma + 1)^2},
\label{eq-fwrap}
\end{equation}
with $N^{\rm contact}_j$ being the number of membrane beads of type $j$ whose centre-of-mass distance to the particle centre is less than $R_{\rm p} + \sigma$. The total wrapping is $w=\sum_{j}w_j$. Since we sometimes observe uptake of non-completely wrapped nanoparticles ($w < 1$), we chose to consider a nanoparticle endocytosed if $w > 0.8$ and its centre-of-mass is located below the fully healed mother membrane. The length of each simulation was $10^7$ steps with a time step of 0.008$\tau$, where $\tau$ is the unit of time. %~\footnote{Changing the time step to $dt=0.005$ or changing the friction coefficient did not affect the thermodynamic results.}.
All the simulations were run with our implementation of the membrane model into the molecular dynamics package LAMMPS~\cite{plimpton1995fast}.

\clearpage
\setcounter{figure}{0}    
\setcounter{equation}{0}
\setcounter{section}{0}
\begin{widetext}
\renewcommand{\part}{}
\begin{center}
{\Large \textbf{Supplementary Information: Controlling cargo trafficking in multicomponent membranes}}\\
\end{center}
\vspace{1cm}

\renewcommand{\theequation}{S\arabic{equation}}
\renewcommand{\thepage}{S\arabic{page}}
\renewcommand{\thefigure}{S\arabic{figure}}
\setcounter{page}{1}

This Supplementary Information provides details of the theoretical derivation of the mean field model (Section I), the simulation procedures (Section II) and thermodynamic integration (Section III), and supporting results and figures (Section IV). 
%%%%%%%%%%%%%%%%%%%   Theory  %%%%%%%%%%%%%%%%%%%%%%%%
\section{Analytical model derivation}
We consider a fluid membrane discretised into small patches of size $a^2 \approx 25nm^2$. Each patch represents one component. Different component types are denoted by an index $_j$ and the membrane is defined by a unit composition vector $\ff = [f_1, f_2, f_3, ...]$ specifying the fractions of different component types within the membrane. Components can represent membrane lipids, embedded transmembrane receptors, or inert membrane proteins. Each component type also has a spontaneous curvature $c_{0,j}$ and (mean and Gaussian) bending rigidity moduli $\kappa_j$, $\bar{\kappa}_j$ associated with it. 

\subsection{Curvature free energy}
The theory is based on a Helfrich Hamiltonian for the curvature free energy density:
\begin{equation}
H_c = \frac{\kappa}{2} (2C - c_{0})^2 + \bar{\kappa} K \;,
\label{eq-fc}
\end{equation}
with $C$ and $K$ the local mean and Gaussian curvature of the membrane, and $c_0$ the local preferred curvature. The total elastic free energy of a membrane of size $A$ is usually obtained by integrating Eq. \eqref{eq-fc}, 
\begin{equation}
F_c = \int H_c\, dA\;.
\label{eq-fcint}
\end{equation}
Such integral form implies a strong assumption of locality: all variables pertinent to the elastic free energy are local. The integral form is valid as a macroscopic (thermodynamic) descriptor of an elastic membrane. On a microscopic level however, the atomistic nature implies that the above integral should be computed as a sum over individual discreet building blocks. 

The lipid bilayer membrane is made of finite sized components -- lipids. The thickness of the bilayer is around 5nm, therefore, the smallest length scale, where the local picture of the Helfrich integral is expected to apply, is of the order of membrane thickness $a= 5$nm. We discretise the Hefrich integral into a sum over $N$ components of size $a^2$  
\begin{equation}
F_c = \sum_{j'}^{N} H_{c,j'}  \, a^2 \;,
\label{eq-Fc1sum}
\end{equation}
with $H_{c,j'}$ the curvature free energy density of component $j'$, Eq. \eqref{eq-fc}. An implicit assumption remains: neighbouring components in the membrane are independent. We have not specified the nature of the component, it could represent a lipid patch, a protein surrounded with a few lipids or a protein cluster, for example. The component size could also be larger than 5nm, however, in that case the component does not directly map to a transmembrane protein, but a multitude of proteins and lipids. Choosing the component size at the lower limit (5nm) ensures that each component contains at most a few closely packed proteins; lateral extension of a transmembrane proteins typically being around 2nm. For simplicity we use a picture where each component represents either pure lipids or a single transmembrane protein (surrounded by a few lipids), but the theory can be applied also to multi-protein components.

%%%%%%%%%%%%%%%%%%%   MF   %%%%%%%%%%%%%%%%%%%%%%%%%%%
\subsection{Mean field}
In the system under study the membrane is either flat or wrapped around a particle. In the following we assume a constant mean curvature approximation: the mean curvature of each component of the membrane is a constant and fully determined by the shell radius: $C_j=-1/R_{\rm w}$. Equivalently for the flat membrane the mean curvature of all components is zero: $C=0$. This approximation amounts to neglecting correlations between local mean curvature and spontaneous curvature of the membrane. For example a stoichiometric mixture of components with equal positive and negative spontaneous curvatures in a flat membrane would yield an overall flat membrane of zero mean curvature. However, the mean-field formalism constrains every component of a flat membrane to a zero mean curvature and the free energy is overstated. In the following, all bending moduli are assumed to be the same: $\kappa=\kappa_j$ and  $\bar{\kappa}=\bar{\kappa}_j$. If the bending moduli are not the same the mean curvature $C_j$ would not only be correlated to the spontaneous curvature $c_{0,j}$, but also to the bending rigidity $\kappa_j$, which would lead to further deviation of the optimal membrane shape from the perfectly spherical shell assumed in the mean-field picture.  

The membrane contains a mixture of multiple types of components. Because all components are assumed to have the same imposed mean and Gaussian curvature, $C_{j'} = C$ and $K_{j'}=K$, Eq. \eqref{eq-Fc1sum} can be equally written as a sum over $n$ component types $j$ 
\begin{equation}
F_c = \sum_{j}^{n} N_j H_{c,j} \; a^2  =  N \sum_{j}^{n} f_j  H_{c,j} \; a^2 \;,
\label{eq-Fc2}
\end{equation}
with $N_j$ the number of components of type $j$ present in the membrane, $N=\sum_j N_j$ the total number of components and the component composition fractions $f_j = N_j/N$. We use the notation where prime indices $j'$ refer to a specific component located in the membrane, while normal indices $j$ denote a component type.
The elastic mean-field free energy of a membrane size $A$ with curvature $C$, using Eqs. \eqref{eq-fc} and \eqref{eq-Fc2}, is
\begin{equation}
F_c(A, C, K) = A \sum_{j}^{n} f_j  \left[ \frac{\kappa}{2} (2C - c_{0,j})^2 + \bar{\kappa} K \right] \;.
\label{eq-Fc}
\end{equation}

\subsection{Mean field 2: alternative approach.}
An alternative approach would relax the requirement that the membrane mean curvature is imposed on every component. Instead, the mean curvature is imposed only "on average" over all components in the membrane. This limit assumes no penalty to spatial mean curvature fluctuations: Eqs. \eqref{eq-fc} and \eqref{eq-fcint} are applied directly to a total membrane patch area $A$. This is equivalent to Eq. \eqref{eq-Fc2} with a single large "component" of size  $A$. The spontaneous curvature of the total patch is defined as the average over individual component spontaneous curvatures 
\begin{equation}
\langle c_{0} \rangle =   \frac{1}{N}\sum_{j'}^N c_{0,j'} = \sum_{j}^{n} f_{j} c_{0,j} \;,
\label{eq-avm}
\end{equation}
with $f_j$ the membrane composition vector.  Similarly for the mean square spontaneous curvature: $
\langle c_{0}^2 \rangle =   \sum_{j}^{n} f_{j} c_{0,j}^2 $.
The Hellfrich expression \eqref{eq-Fc1sum} using the mean spontaneous curvature becomes
\begin{equation}
\tilde{F}_c(A,C,K) / A =  \frac{\kappa}{2}  (2C - \langle c_0 \rangle)^2  + \bar{\kappa} K  =  \frac{\kappa}{2} \left( 4C^2 -4 C \langle c_0 \rangle + \langle c_0 \rangle^2 \right) + \bar{\kappa} K   \;.
\label{eq-Fcmean2}
\end{equation}

We compare the above result to the previous mean-field expression (Eq. \eqref{eq-Fc}) applied to a membrane area $A$:
\begin{equation} 
F_c(A, C, K)/A =  \sum_{j}^{n} f_j  \left[ \frac{\kappa}{2} (4C^2 - 4 C c_{0,j} + c_{0,j}^2 ) + \bar{\kappa} K \right] = \frac{\kappa}{2}  \left( 4C^2 -4C \langle c_0 \rangle + \langle c_{0}^2 \rangle \right) +  \bar{\kappa} K \;,
\label{eq-Fcmean1}
\end{equation}
These two free energy expressions are related through a fluctuation term 
\begin{equation}
\tilde{F}_c(A,C,K) = F_c(A,C,K) - A\frac{\kappa}{2} \left( {\langle c_0 \rangle^2  - \langle c_0^2 \rangle} \right) \;.
\label{eq-Fcmf12}
\end{equation}
Importantly, the above fluctuation term refers to spatial fluctuations in the spontaneous curvature of membrane components. It does not refer to mean curvature of the membrane; thermal fluctuations of the local membrane bending are implicitly included in the starting Helfrich expression, Eq. \eqref{eq-fc}.

The two mean-field expression, Eqs. \eqref{eq-Fcmean2} and \eqref{eq-Fcmean1}, conveniently provide a lower and upper bound to the free energy of the membrane patch of size $A$ with some average imposed mean and Gaussian curvature $C$ and $K$. The upper bound \eqref{eq-Fcmean1} is reached in the limit of an infinite lateral membrane tension (curvature imposed on every component), while the lower bound  \eqref{eq-Fcmean1} is realised in the case of an 1D membrane (a chain of components) with zero lateral tension (total curvature of the membrane imposed only at the boundary). For a 2D membrane with zero lateral tension the geometric constraints on bending likely prevent the lower bound of being reached. Entropy of partial ordering of components affects the 2D membrane, but not the 1D chain, because any permutation of components along a chain will result in the same curvature energy (unless the curvature is so large that the chain folds onto itself).  

In the following the upper estimate for the mean-field free energy (Eq. \eqref{eq-Fc}) will be used for all theoretical predictions. The reason for this choice is that the nanoparticle binds to the membrane via receptor attachments and the membrane envelope will likely be tightly wrapped around the particle. 

%%%%%%%%%%%%%%%    MIXING  ENTROPY     %%%%%%%%%%%%%%%%%%%%%%%%%%%%%
%\subsection{mixing entropy of beads} 

%%%%%%%%%%%%%%%%%%   ENDOCYTOSIS FREE ENERGY  %%%%%%%%%%%%%%%%%%%%%
\subsection{Endocytosis free energy}
We attempt to analytically calculate the free energy change upon a particle endocytosis. As stated above, the membrane is composed of different component types with a component vector $\ff$ and spontaneous curvature vector ${\mathbf c}_0$ specifying the spontaneous curvature of all component types. Additionally each component type also has an interaction with a particle captured by the vector $\eeps$. We assume a simple square well interaction potential where $\eeps$ specifies the well depth. The lateral tension and density $\rho$ of the membrane is assumed to be constant. All three vectors ($\ff, {\mathbf c}_0, \eeps$) are of length $n$ with $n$ the number of distinct components in the membrane.

The following notation is used:
\begin{itemize}
\item e.g. $A, N, \ff$ -- standard letters refer to the initial flat membrane {\em before} endocytosis
\item e.g. $\tilde{A}, \tilde{N}, \tilde{\ff}$ -- letters with a tilde  refer to the membrane that remains flat {\em after} the full endocytosis has taken place
\item e.g. $A_{\rm w}, N_{\rm w}, \ff^{\rm w}$ -- letters with $_w$ script  refer to the membrane wrapped around the particle {\em after} the full endocytosis has taken place.
\end{itemize}

Initially, the membrane is flat with a total area $A$ and $\ff$ is the component composition in the membrane. Upon particle endocytosis a small membrane area $A_{\rm w}=4\pi R_{\rm w}^2$ is wrapped around a particle with a membrane shell radius $R_{\rm w}$. The remaining membrane area $\tilde{A}=A-A_{\rm w}$ remains flat. The composition of the wrapped part $\ff^{\rm w}$ is in general different from the remaining flat membrane $\tilde{\ff}$. 
Therefore, the two conservation laws are: Area conservation
\begin{equation}
A = \tilde{A} + A_{\rm w} 
\label{eq-cons1}
\end{equation}
and component number conservation: $N=\tilde{N} + N_{\rm w}$, which applies to every component type, hence, 
\begin{equation}
\ff A = \tilde{\ff} \tilde{A} + \ff^{\rm w} A_{\rm w}  \;.
\label{eq-cons2}
\end{equation}
We have assumed that the lateral membrane tension and 2D component density $\rho = \frac{N}{A}=\frac{\tilde{N}}{\tilde{A}}=\frac{N_{\rm w}}{A_{\rm w}} = const.$ remains unchanged during the endocytosis. 

Furthermore, we assume that lateral component diffusion is fast compared to the endocytosis timescale. Under these conditions the components can be treated as independently adsorbing to the particle. The process of endocytosis is rather complicated, however, we are only interested in the free energy change between the initial and final state. We assume that during all of the endocytosis process the (partially) wrapped part of the membrane is in contact with the remainder of the membrane (the reservoir).  Therefore, the wrapped components composition is given by the generalised Langmuir adsorption isotherm:
\begin{equation}
f^{\rm w}_j = \frac{e^{\beta (\tilde{\mu}_j - F_{c,j}(R_{\rm w}) - \epsilon_j)}}{\sum_{j} e^{\beta (\tilde{\mu}_j - F_{c,j}(R_{\rm w})- \epsilon_j)} } \;
\end{equation}
with
\begin{equation}
F_{c,j}(R_{\rm w}) = F_c \frac{a^2}{A}  = a^2 \sum_{j}^{n} f_j  \left[ \frac{\kappa}{2} (2/R_{\rm w} + c_{0,j})^2 + \bar{\kappa} /R_{\rm w}^2 \right] \;
\label{eq-fcpc}
\end{equation}
the curvature free energy \eqref{eq-Fc} per single component of lateral size $a^2$. $\epsilon_j$ is the component-particle interaction energy and $\tilde{\mu}_j$ is the chemical potential of component type $j$ in the flat membrane area $\tilde{A}$. The chemical potential consists of the ideal contribution (logarithm of the density) and the excess part which contains the curvature free energy of a component embedded in a flat membrane. In general there are other contribution to the excess chemical potential, such as component interactions interactions and membrane tension contribution. However, we assume that all other contributions remain constant upon particle wrapping, and only the curvature free energy changes. 

The chemical potential is 
\begin{equation}
\tilde{\mu}_j = k_{\rm B}T \ln (\tilde{f}_j) + f_{c,j}^{0}\;,
\end{equation}
with
$f_{c,j}^0 \equiv f_{c,j}(\infty)$ the curvature free energy in a flat membrane ($R_{\rm w}\to \infty$). The adsorbed composition can, therefore, be written as
\begin{equation}
f^{\rm w}_j = \frac{\tilde{f}_j \, e^{\beta (f_{c,j}^{0} - f_{c,j}(R_{\rm w}) - \epsilon_j)}}{\sum_{j} \tilde{f}_j \, e^{\beta (f_{c,j}^0 - f_{c,j}(R_{\rm w}) - \epsilon_j)} } = \tilde{f}_j \tilde{K}_j \;.
\label{eq-cw}
\end{equation}
where in the last step we have defined an equilibrium constant 
\begin{equation}
\tilde{K}_j = \frac{e^{\beta (f_{c,j}^{0} - f_{c,j}(R_{\rm w}) - \epsilon_j)}}{\sum_{j} \tilde{f}_j \, e^{\beta (f_{c,j}^0 - f_{c,j}(R_{\rm w}) - \epsilon_j)} }
\label{eq-tK}
\end{equation}
 specifying how strong individual component types adsorb to the particle. The equilibrium constant $\tilde{K}_j $ is not to be confused with the Gaussian curvature $K$. Another conservation relation emerges: $\sum_j \tilde{f}_j \tilde{K}_j = 1$.  Using Eqs. (\ref{eq-cons1}, \ref{eq-cons2}, \ref{eq-cw}) we can write the composition of the remaining membrane as:
\begin{equation}
\tilde{f}_j  = \frac{f_j}{1+\frac{A_{\rm w}}{A}(\tilde{K}_j - 1)} \;.
\label{eq-ct}
\end{equation}

The total free energy change upon a particle endocytosis can be written in terms of individual contributions due to membrane curvature, binding to the particle, mixing entropy and lateral tension $\Pi$: 
\begin{equation}
\Delta F  =  \Delta F_c + \Delta \epsilon  - T \Delta S -  \Pi A_{\rm w} \;.
\label{eq-delF}
\end{equation}
The change in the curvature free energy upon endocytosis is obtained by the free energy of the wrapped membrane shell and the remaining flat membrane, minus the initial state which is a flat membrane of area A
\begin{equation}
\Delta F_c =  F^{\rm w}_c + \tilde{F}_c - F_c  = A_{\rm w} \frac{\kappa}{2}  \sum_j f^{\rm w}_j (2/R_{\rm w} + c_{0,j})^2  + \tilde{A} \frac{\kappa}{2}  \sum_j  \tilde{f}_j (c_{0,j})^2  - A \frac{\kappa}{2}  \sum_j f_j (c_{0,j})^2  + 4 \pi \tilde{\kappa} \;.
\label{eq-delFc}
\end{equation}
This expression was obtained using Eq. \eqref{eq-Fc} with the mean curvature determined by the shell radius $C=-1/R_{\rm w}$ for an endocytosed shell, and $C=0$ for a flat membrane. The contribution of the Gaussian curvature is $4 \pi \tilde{\kappa}$ due to Gauss-Bonnet theorem. 

The change in binding interaction energy between membrane components and the particle is trivial:
\begin{equation}
\Delta \epsilon =\frac{A_{\rm w}}{\Delta a} \sum_j f^{\rm w}_j \epsilon_j
\label{eq-delFe}
\end{equation}
because the component-particle interaction vector $\eeps$ is a constant and interaction is only present in the final fully wrapped endocytosed state.

Lastly, the membrane is assumed to be in a fluid state. The mixing entropy of components is given by the Gibbs expression for the entropy per component: 
$s = - k_{\rm B} \sum_j f_j \ln(f_j)$,
where we remember $f_j$ as the fraction of component $j$ in the membrane and $k_{\rm B}$ is the Boltzmann constant. $f_j$ can also be seen as the probability that a randomly chosen component is of type $j$.
The difference in entropy of mixing of different component types upon endocytosis is
\begin{equation}
\Delta S = S^{\rm w} + \tilde{S} - S  = -\frac{k_{\rm B}}{a^2} \left[  A_{\rm w} \sum_j f^{\rm w}_j \ln f^{\rm w}_j  + \tilde{A} \sum_j \tilde{f}_j \ln \tilde{f}_j - A \sum_j f_j \ln f_j \right] 
\label{eq-delFs}
\end{equation}
using the Gibbs entropy per single component and entropy extensivity $S=\frac{A}{a^2} s$. 

These relations Eqs. (\ref{eq-delF}-\ref{eq-delFs})  result in a closed form expression for the endocytosis free energy as a function of the membrane composition $\ff$, curvature vector ${\mathbf c}_0$, interaction vector $\eeps$ and the wrapped membrane shell radius $R_{\rm w}$:
\begin{equation}
\Delta F (\ff,{\mathbf c}_0,\eeps,R_{\rm w}) = A_{\rm w} \sum_j f_j K_j \left[ \frac{\epsilon_j}{a^2} + \frac{2\kappa}{R_{\rm w}}\left(\frac{1}{R_{\rm w}} + c_{0,j}\right) + \frac{k_{\rm B}T}{a^2} \ln(K_j) \right]  - A_{\rm w} \Pi  + 4 \pi \bar{\kappa} + A_{\rm w} \mathcal{O}(A_{\rm w} / A )\;.
\label{eq-delFfin}
\end{equation}
The first term inside the square brackets captures the binding of components to the particle, the second terms curvature mismatch penalty, and the third term the effect of the membrane composition change between the flat membrane and the wrapped part with the equilibrium constant
\begin{equation}
K_j = \frac{ e^{-\beta \left[ \epsilon_j+ \frac{2\kappa a^2}{R_{\rm w}}(1/R_{\rm w}+ c_{0,j}) \right] }}{\sum_{j} f_j \, e^{-\beta \left[ \epsilon_j+ \frac{2\kappa a^2}{R_{\rm w}}(1/R_{\rm w}+ c_{0,j}) \right] }} \;.
\end{equation}
 The pre-factor $A_{\rm w} = 4 \pi R_{\rm w}^2$ is the wrapped membrane area and $a^2$ the individual component lateral size.  Finally, the last term $\mathcal{O}(A_{\rm w} / A )$ captures all terms of order $A_{\rm w}/A$ and higher. For a large membrane $A_{\rm w} \ll A$ these terms can be neglected. 
 
 %%%%%%%%%%%%%%%%    ANALYTICAL  DERIVATION   %%%%%%%%%%%%%%%%%%%%%%%%%%%%  
 \subsection{Detailed derivation of endocytosis free energy}
 Here we provide a step by step derivation procedure of the endocytosis free energy Eq. \eqref{eq-delFfin} from Eqs. (\ref{eq-delF}-\ref{eq-delFs}).

Firstly, we focus on the curvature free energy change Eq. \eqref{eq-delFc}.
Inserting Eqs. \eqref{eq-cw} and \eqref{eq-ct} and $\tilde{A} = A - A_{\rm w}$, rearranging and canceling out a few terms we find
\begin{equation} 
 \Delta F_c  = 2 A_{\rm w} \kappa \sum_j \frac{f_j\tK_j}{1+\frac{A_{\rm w}}{A} (\tilde{K}_j -1)} \left( 1/R_{\rm w}^2 + c_{0,j}/R_{\rm w}\right) \;.
 \label{eq-delFc2}
\end{equation}
The interaction energy contribution Eq. \eqref{eq-delFe} is slightly rewritten by inserting Eqs. \eqref{eq-cw} and \eqref{eq-ct}: 
\begin{equation}
\Delta \epsilon =\frac{A_{\rm w}}{a^2} \sum_j  \epsilon_j \frac{f_j\tK_j}{1+\frac{A_{\rm w}}{A} (\tilde{K}_j -1)}
\label{eq-delFe2}
\end{equation}
Finally, the entropy change Eq. \eqref{eq-delFs} is also rewritten by inserting Eqs. \eqref{eq-cw} and \eqref{eq-ct} and $\tilde{A} = A - A_{\rm w}$: 
\begin{equation}
\frac{\Delta S}{k_{\rm B}}= -\frac{A_{\rm w}}{a^2} \sum_j \frac{f_j}{1+\frac{A_{\rm w}}{A}(\tK_j - 1)} \left[ \tK_j \ln \tK_j + \left(1-\frac{A}{A_{\rm w}} - \tK_j \right) \ln \left(1+\frac{A_{\rm w}}{A}(\tK_j-1)\right)\right]
\label{eq-delFs2}
\end{equation}

We now take the limit of a large membrane: $A_{\rm w} / A \to 0$. This implies: $\tilde{c}_j \to c_{j}$ (from Eq. \eqref{eq-ct}) and $\tK_j \to K_j$ (from Eq. \eqref{eq-tK}) where the equilibrium constant $K_j$ is defined as:
\begin{equation}
K_j = \frac{ e^{\beta (f_{c,j}^{0} - f_{c,j}(R_{\rm w})- \epsilon_j)}}{\sum_{j} f_j \, e^{\beta (f_{c,j}^0 - f_{c,j}(R_{\rm w})- \epsilon_j)} } = \frac{ e^{-\beta \left[ \epsilon_j+ \frac{2\kappa a^2}{R_{\rm w}}(1/R_{\rm w}+ c_{0,j}) \right] }}{\sum_{j} f_j \, e^{-\beta \left[ \epsilon_j+ \frac{2\kappa a^2}{R_{\rm w}}(1/R_{\rm w}+ c_{0,j}) \right] }} \;.
\label{eq-K}
\end{equation}
In the second step in the above equation the Helfrich free energy per component (Eq. \eqref{eq-fcpc}) was inserted.
Taking the limit $A_{\rm w} / A \to 0$ the curvature free energy \eqref{eq-delFc2} becomes:
\begin{equation} 
 \Delta F_c  = 2 A_{\rm w} \kappa \sum_j f_j K_j \left( 1/R_{\rm w}^2 + c_{0,j}/R_{\rm w}\right)  + A_{\rm w}\mathcal{O}(A_{\rm w}/A)\;.
 \label{eq-delFc3}
\end{equation}
The interaction energy \eqref{eq-delFe2}:
\begin{equation}
\Delta \epsilon = \frac{A_{\rm w}}{a^2} \sum_j  \epsilon_j f_j K_j + A_{\rm w}\mathcal{O}(A_{\rm w}/A)\;.
\label{eq-delFe3}
\end{equation}
and the entropy \eqref{eq-delFs2} can be simplified by expanding the logarithm to the first order and using equalities $\sum_{j} f_j=1$ and $\sum_{j} f_j K_j=1$ :
\begin{equation}
\frac{\Delta S}{k_B} = -\frac{A_{\rm w}}{a^2} \sum_j f_j K_j \ln K_j +  A_{\rm w}\mathcal{O}(A_{\rm w}/A)\;.
\label{eq-delFs3}
\end{equation}
$\mathcal{O}(A_{\rm w} / A )$ captures all terms of order $A_{\rm w}/A$ and higher. 
Summing up the individual contributions (Eq. \eqref{eq-delF}) the final result (Eq. \eqref{eq-delFfin}) follows.

%%%%%%%%%%%%%%%    SIMULATION   MODEL

\section{Simulation model}
\begin{figure*}[ht!]
\center
\includegraphics[width=0.8\textwidth]{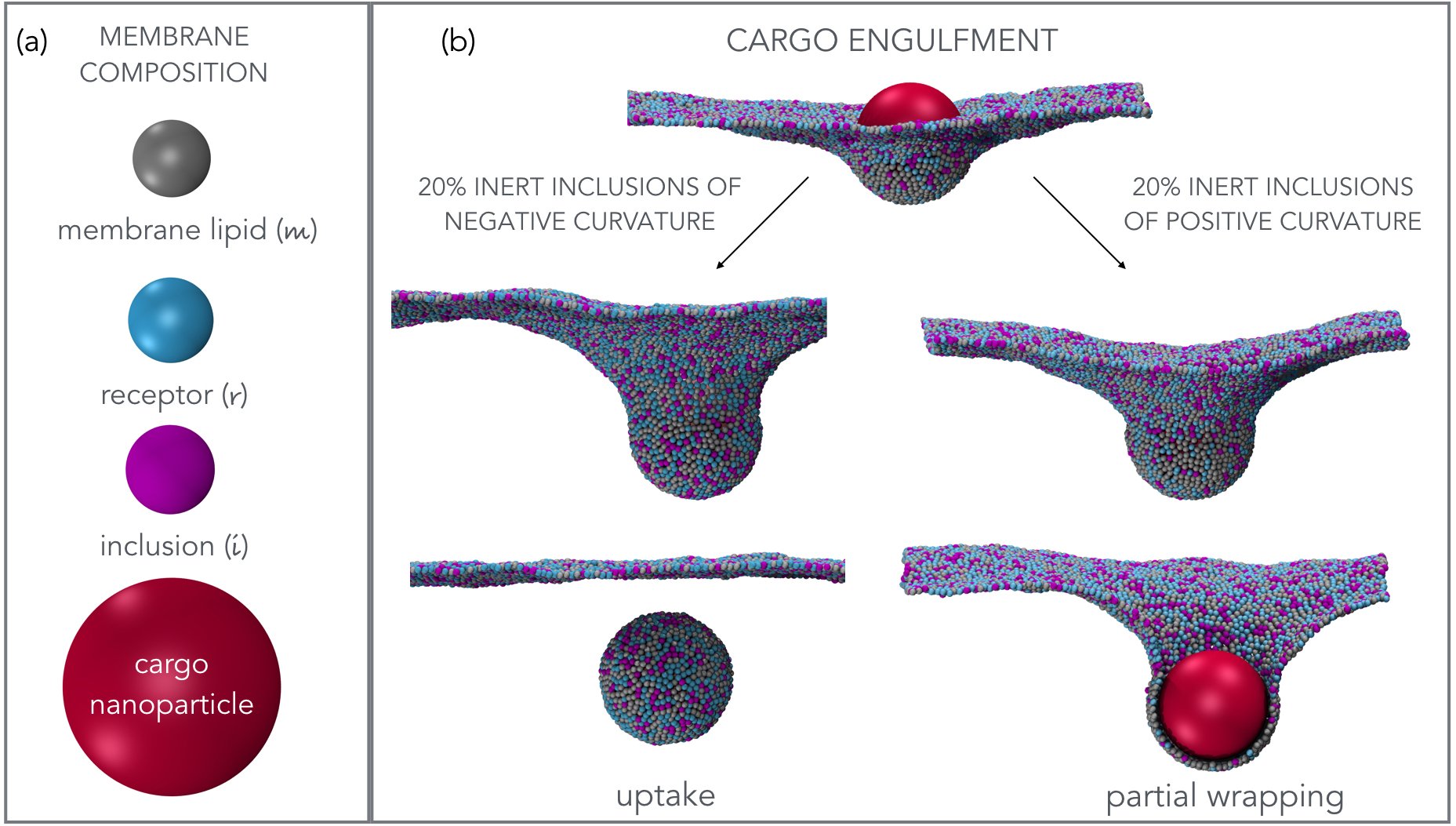}
\caption{\textbf{Simulation model}. (a) The membrane is composed of three types of beads: non-binding lipid beads ('m') of zero spontaneous curvature, cargo-binding receptor beads ('r') of spontaneous curvature $c_{\rm 0,r}$, and inert (non-binding) inclusions ('i') of spontaneous curvature $c_{\rm 0,i}$ ~\cite{yuan2010one}. The cargo (modelled as generic nanoparticle) is much bigger than either of the membrane beads. (b) Cargo binds to membrane receptors, deforms the membrane, and in case of 20\% of inert inclusions of negative curvature becomes completely wrapped by the membrane and buds off, while in the case of 20\% of inert inclusions of positive curvature it stays partially wrapped without being endocytosed within the time of the simulation of $10^7$ time steps.}
\label{fig-sim}
\end{figure*}

The membrane is modelled using a coarse grained one-particle thick model~\cite{yuan2010one}, which we implemented in the LAMMPS MD package. Following the notation from the original paper, we choose the parameters $\epsilon_{\rm bead-bead}=4.34k_{\rm B}T, \xi=4, \mu=3, r_{\rm cut}^{\rm bead-bead}=1.12$ for all the membrane components. The repulsion potential between the nanonanoparticle and membrane beads is described by the Weeks-Chandler-Anderson potential

\begin{equation}
U_{\rm WCA}(r) = \epsilon_{\rm WCA} \left[ 1+ 4 \left(\frac{\sigma}{r - R_{\rm p}} \right)^{12} - 4 \left(\frac{\sigma}{r - R_{\rm p}} \right)^{6}\right] \;,
\end{equation} 
for $0 \le r - R_{\rm p} \le 2^{1/6}\sigma$ and $U_{\rm WCA}(r)=0$ otherwise. $r$ is the bead to particle centre-of-mass distance, $R_{\rm p}$ and $\sigma$ are the nanoparticle and bead radius, respectively. We chose the interaction strength $\epsilon_{\rm WCA} = \epsilon_{\rm bead-bead}$. 

The interaction potential between nanoparticle and `receptor' membrane beads is modelled as a cut-and-shifted Lennard Jones potential
\begin{equation}
U_{\rm bead-particle}(r) = 4\epsilon^* \left[ \left(\frac{\sigma}{r - R_{\rm p}} \right)^{12} - \left(\frac{\sigma}{r - R_{\rm p}} \right)^{6}\right]  + U_{cs} \;,
\end{equation} 
for $ r - R_{\rm p} \le 2.6\sigma $ and $0$ otherwise. $U_{cs}=-U_{\rm bead-particle}(2.6\sigma)$.
This potential is implemented in the LAMMPS MD package as "lj/expand".

 %%%%%%%%%%   Gaussian rigidity  %%%%%%%%%%%%%%
\section{Determining Gaussian bending stiffness}

The standard bending rigidity $\kappa$ can be obtained from the fluctuation spectrum of the membrane.  the Gaussian bending rigidity $\bar{\kappa}$, however, is trickier to calculate. We obtain the Gaussian bending rigidity of the membrane model both by theoretical considerations and by performing thermodynamic integration. Both approaches are presented below, theoretical considerations yield $\bar{\kappa} = -\kappa$ and thermodynamic integration $\bar{\kappa} \approx -\kappa \approx - 22k_{\rm B}T$ which also agrees with the value obtained from the membrane fluctuation spectrum~\cite{yuan2010one}. We, therefore, use $\bar{\kappa} = -\kappa = 22k_{\rm B}T$ when comparing analytical and simulation results.

\subsection{Thermodynamic integration}
We performed Monte Carlo simulations with thermodynamic integration scheme to calculate the mean $\kappa$ and Gaussian $\bar{\kappa}$ bending rigidity of the membrane. Three sets of simulations with different membrane topologies were performed: flat membrane (periodic boundary in lateral directions), cylindrical membrane (periodic boundary in the cylinder axis) and spherical membrane. $N=5000$ beads were used for all simulations and no applied external pressure, $5\cdot 10^5$ MC cycles were used for all simulations, unless noted otherwise. The radius of the cylindrical membrane was half the radius of a spherical membrane $R_{\rm c} = R_{\rm s} / 2$. Such a choice allows us to directly determine the bending rigidities from the free energies of the three different membrane configurations
\begin{equation}
\kappa = \frac{F_{\rm c} - F_{\rm f}  }{8 \pi}  \;, \quad \quad
\bar{\kappa} = \frac{F_{\rm s} - F_{\rm c}  }{4 \pi} \;,
\end{equation}
where $F_{\rm f}$, $F_{\rm c}$ and $F_{\rm s}$ are the free energies of the membrane in the flat, cylindrical and spherical configuration, respectively.

The tensionless flat membrane 2D density was measured to be $\rho_{\rm beads} = 1.206$, therefore we chose the spherical membrane radius as $R_{\rm s} = \sqrt{\frac{N}{4\pi \rho}}$ and the cylindrical membrane radius  $R_{\rm c} = R_{\rm s} / 2 $, with the cylinder height chosen to conserve the number of beads $h_{z, \rm c} = \frac{N}{2\pi R_{\rm c} \rho_{\rm beads} } $. 

The reference state of the thermodynamic integration is a 2D ideal gas with density $\rho_{\rm beads} = 1.206$ confined to lie on a flat, cylindrical or spherical surface. The free energy of the thermodynamic integration proceeded in two steps. First the bead-bead interaction potential strength was increased from 0 (ideal gas) to $\epsilon_{\rm bead-bead} = 4.34k_{\rm B}T$. A total of 201 simulations were performed for each topology with the interaction parameter linearly spaced $\epsilon_{i, \rm bead-bead} = \epsilon_{\rm bead-bead} \frac{i}{200} $. The first calculation ($i=0$) was performed at $\epsilon_{0, \rm bead-bead} =0.001k_{\rm B}T$.  The results of these simulations are shown on Figure~\ref{fig-thermoint}{\bf a)}. The integration was performed on the energy differences between different membrane topologies which avoids the problem of integrand divergence at $\epsilon \sim 0$. The thermodynamic integration using Simpson's rule yields bending rigidities $\kappa_{\rm 2D} = 23.2k_{\rm B}T$ and $\bar{\kappa}_{\rm 2D} = -25.5 k_{\rm B}T$. Note that the membrane beads were confined to a 2D surface.

\begin{figure*}[ht!]
\center
\subfigure[]{\includegraphics[width=0.50\textwidth]{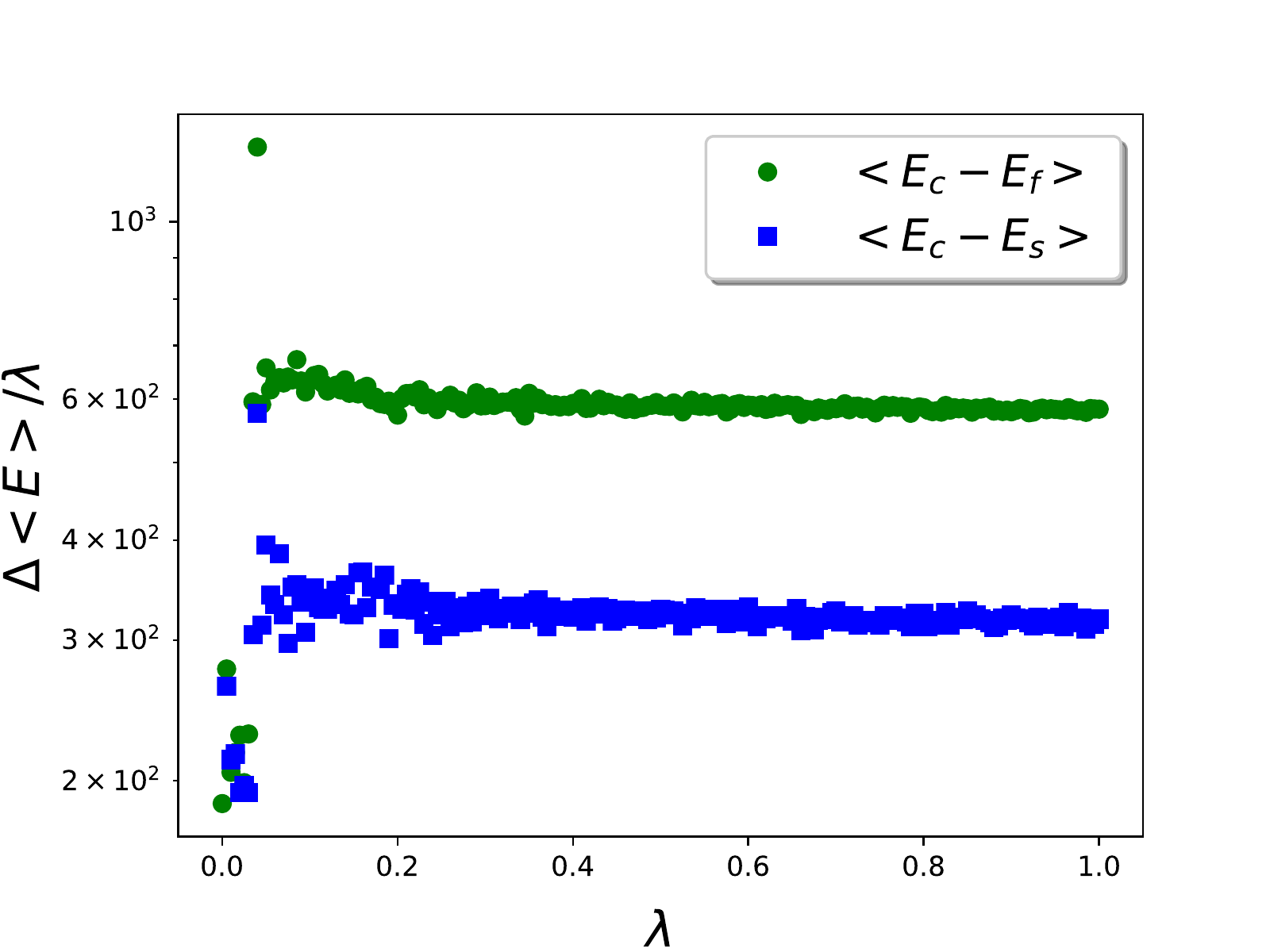}} \ \
\subfigure[]{\includegraphics[width=0.45\textwidth]{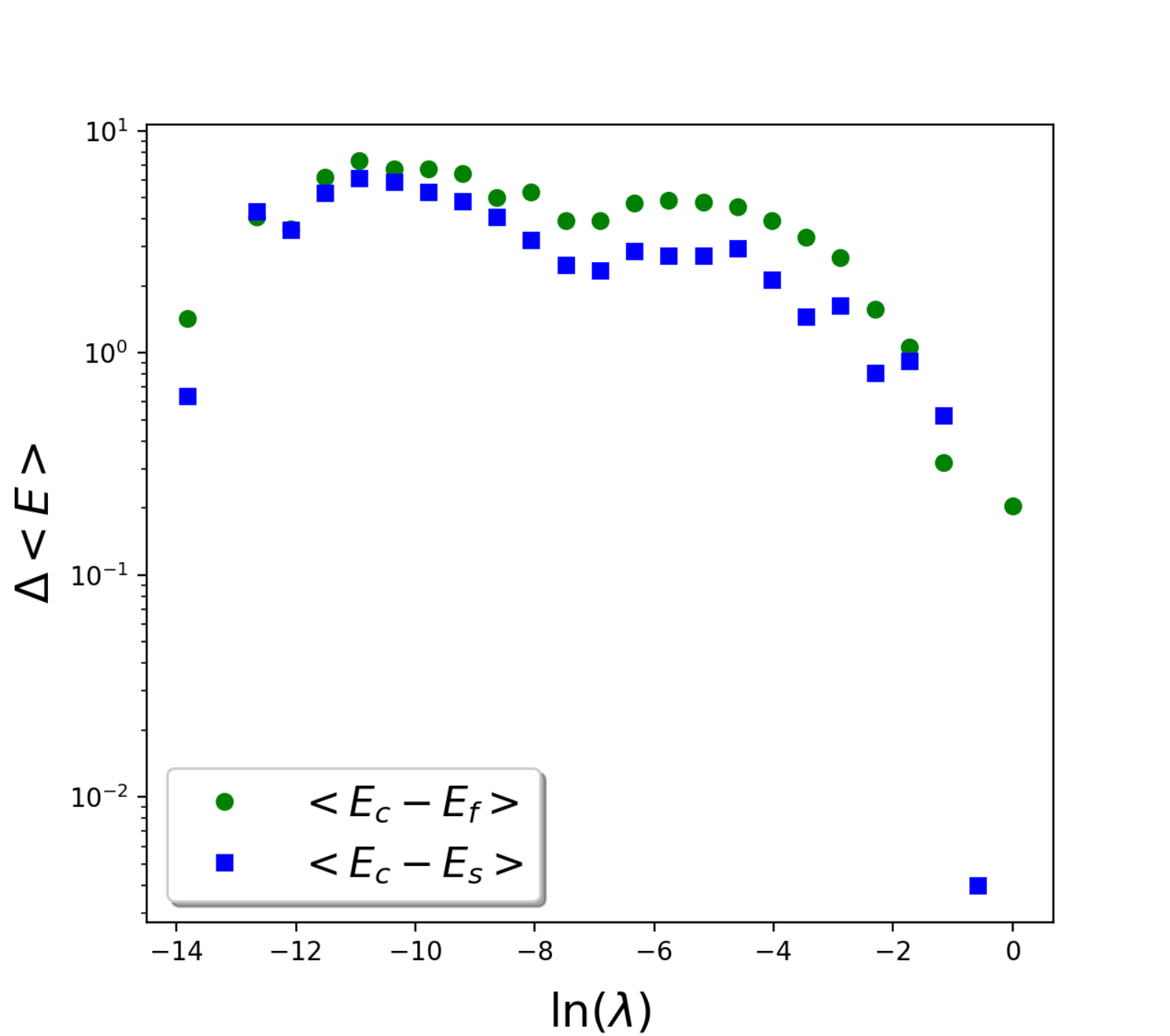}}
\caption{Thermodynamic integration results. Average potential energy difference between different membrane topologies as a function of the thermodynamic integration parameter $\lambda$. $E_f, E_c$ and $E_s$ denote the potential energies of the three distinct topologies (flat, cylindrical, spherical).  (a) The first thermodynamic integration step: changing the bead-bead interaction $\lambda = \epsilon_{i, \rm bead-bead} /  \epsilon_{\rm bead-bead}$, and (b) The second thermodynamic integration step: changing the harmonic confining potential $\lambda = k_i / k_{\rm max}$.}
\label{fig-thermoint}
\end{figure*}

In the second thermodynamic integration step the membrane beads are relaxed to allow for fluctuations in the direction normal to the membrane. A harmonic confining potential $U(r) = \frac{k}{2}r^2$ is introduced for each bead, with $r$ the normal distance between the bead and the confining surface. The confining surfaces are identical to the surfaces used above (flat, cylinder, sphere). 25 values for the confining potential strength $k$ are logarithmically spaced between $k_{\rm min} = 0.01 k_{\rm B}T/\sigma^2$ and $k_{\rm max}=10000 k_{\rm B}T/\sigma^2$. Thermodynamic integration yields the correction to the bending rigidities: $\kappa_{\rm relax} = -2.0k_{\rm B}T$ and $\bar{\kappa}_{\rm relax} = 3.0 k_{\rm B}T$. 

The bending rigidities of the membrane are therefore: 
\begin{eqnarray}
\kappa &=& \kappa_{2D} + \kappa_{\rm relax} = 21.2 k_{\rm B}T \;, \\
\bar{\kappa} &=& \bar{\kappa}_{2D} + \bar{\kappa}_{\rm relax} = -22.5 k_{\rm B}T  \;.
\end{eqnarray}
The value for $\kappa$ obtained by thermodynamic integration agrees well with the bending rigidity calculated from the fluctuation spectrum $\kappa_{\rm fs}\approx 22 k_{\rm B}T$~\cite{yuan2010one}.

%%%%%%%%%%%%%   Analytical  %%%%%%%%%%%%%%%%%%%%%%%
\subsection{Analytical considerations}

The Gaussian bending rigidity can also be estimated assuming a simple microscopic model of the membrane. The membrane consists of a monolayer of beads. Individual beads are not deformable and have cylindrically symmetry around a director axis. Nearest neighbour beads $i,j$ have a harmonic pair potential with spring constant $k$ for the director bending
\begin{equation}
U_{ij} = \frac{k}{2}(\theta_{ij} - \theta_0)^2 \;,
\end{equation}
where $\theta_{ij}$ is the angle between the directors of the two beads and $\theta_0$ is a constant specifying the preferred orientation between the two beads. Mean distance between nearest neighbours is $d_o$ and each bead has $z$ nearest neighbours. The angle between neighbouring beads is related to the membrane curvature: $\sin (\theta_{ij} / 2) = d_0 / (2R_{ij})$ from which we obtain for small curvatures $\theta_{ij} \approx d_0 / R_{ij} = d_0 C_{ij}$, with $R_{ij}=1/C_{ij}$ the radius of the curved membrane with curvature $C_{ij}$. Therefore, the above equation can be rewritten to
\begin{equation}
U_{ij} = \frac{k d_0^2}{2}(C_{ij} - c_0)^2 \;,
\end{equation}
with the spontaneous curvature $c_0 = \theta_0 / d_0$. 

Using this microscopic model we can calculate the macroscopic bending rigidities $\kappa$ and $\bar{\kappa}$ by comparing the energy of bending obtained from the microscopic model with the Helfrich hamiltonian. For a flat membrane $C_{ij}=0$ the microscopic model yields the membrane elastic energy per bead
\begin{equation}
U^{\rm flat} = \frac{z}{2} \frac{k d_0^2}{2}(c_0)^2 \;.
\label{eq-gcUf}
\end{equation}
The prefactor $\frac{z}{2}$ is simply the number of pair interactions per bead. Cylindrical membrane yields the elastic energy of 
\begin{equation}
U_{\rm bend}^{\rm cylinder} = \frac {z}{4} \frac{k d_0^2}{2}(c_0)^2 + \frac{z}{4} \frac{k d_0^2}{2}(1/R_{\rm c} - c_0)^2 \;
\end{equation}
using a mean-field like approximation where half of the nearest neighbours are parallel $C_{ij} = C_1 = 0$ and the other half are confined to a curvature of $C_{ij}=C_2 = 1/R_{\rm c}$ with $R_{\rm c}$ the cylinder radius and $C_1$, $C_2$ the principal curvatures of the membrane.
Lastly, spherical membrane results in 
\begin{equation}
U_{\rm bend}^{\rm sphere} =  \frac{z}{2} \frac{k d_0^2}{2}(1/R_{\rm s} - c_0)^2 \;
\end{equation}
because all nearest neighbours feel the same bending curvature $C_{ij} = C_1 = C_2 = 1/R_{\rm s}$ and $R_{\rm s}$ is the sphere radius.

The Helfrich expression for the elastic free energy density is
\begin{equation}
H = \frac{\kappa}{2}(2C - c_0)^2 + \bar{\kappa} K \;
\label{eq-gcH}
\end{equation}
where $C = (C_1 + C_2) / 2$ is the mean curvature of the membrane and $K= C_1 C_2$ the Gaussian curvature with $C_1$ and $C_2$ the principal curvatures of the membrane. 
For a flat membrane Helfrich yields $H^{\rm flat} = \frac{\kappa}{2}(c_0)^2$, cylindrical $H^{\rm cylinder} = \frac{\kappa}{2}(1/R_{\rm c} - c_0)^2$ and spherical $H^{\rm sphere} = \frac{\kappa}{2}(2/R_{\rm s} - c_0)^2 + \bar{\kappa}/ R_{\rm s}^2$. 

The elastic energy obtained from the microscopic model must be the same as the Helfrich expression for all three membrane topologies up to a common additive constant. Assuming that the membrane curvature radii considered are always large as compared to the microscopic bead size, $d_0/R \sim 0$, the local configuration of beads will not be affected by the curvature and the number of neighbours per bead $z$ can be treated as a constant. Hence, the only possible relation connecting Eqs. (\ref{eq-gcUf}-\ref{eq-gcH}) is
\begin{equation}
\kappa = -\bar{\kappa} = \frac{z k d_0^2}{4} \;
\end{equation}
and $U = H + \frac{\kappa}{2} c_0^2$. 

Therefore, for a membrane composed of a monolayer of cylindrically symmetric beads and small curvatures the Gaussian bending rigidity is opposite of the mean bending rigidity $\bar{\kappa} = -\kappa$. 
This result is supported by the thermodynamic integration discussed above. We therefore use the value of bending rigidities of $\bar{\kappa} = -\kappa = 22k_{\rm B}T$ for both the mean and Gaussian bending rigidity when comparing analytical and simulation results of nanoparticle endocytosis. 

%%%%  SUPPORTING RESULTS
\section{Supporting results}
\subsection{Membrane stability}
Analytical model can provide insight into membrane stability depending on the composition of curved inclusions.  We consider only the membrane, without cargo, and determine the thermodynamic stability of a flat membrane with respect to budding and formation of a separate vesicle. Eq.~\eqref{eq-delFfin} is solved numerically to obtain inert inclusion curvature as a function of the vesicle radius $R_{\rm w}$: $c_{\rm 0,i} ( R_{\rm w} | \Delta F= 0)$ in the limiting case of zero free energy cost of forming the vesicle. Phase diagrams on Figure~\ref{fig-memstab} show that a flat membrane with zero total spontaneous curvature is thermodynamically stable if the absolute value of spontaneous curvature of individual components is not large. The larger the fraction of curved components $f_{\rm i}$ the lower the limiting value of spontaneous curvature $c_{\rm 0,i}$. On the other hand, flat membranes with non-zero total spontaneous curvature are, as expected, always unstable with respect to a formation of a vesicle.

\begin{figure*}
\center
\subfigure[]{\includegraphics[width=0.45\textwidth]{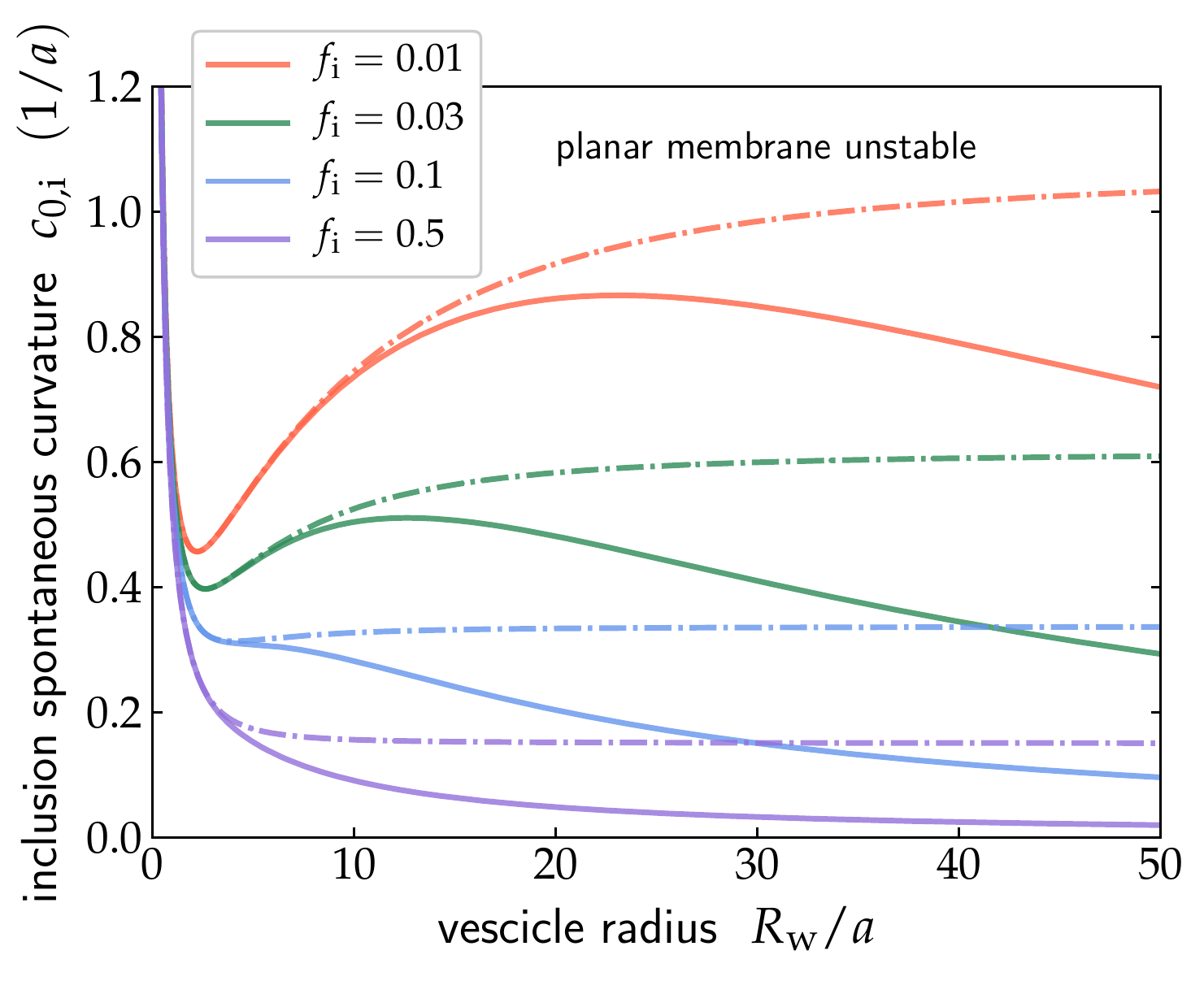}} \ \ 
\subfigure[]{\includegraphics[width=0.45\textwidth]{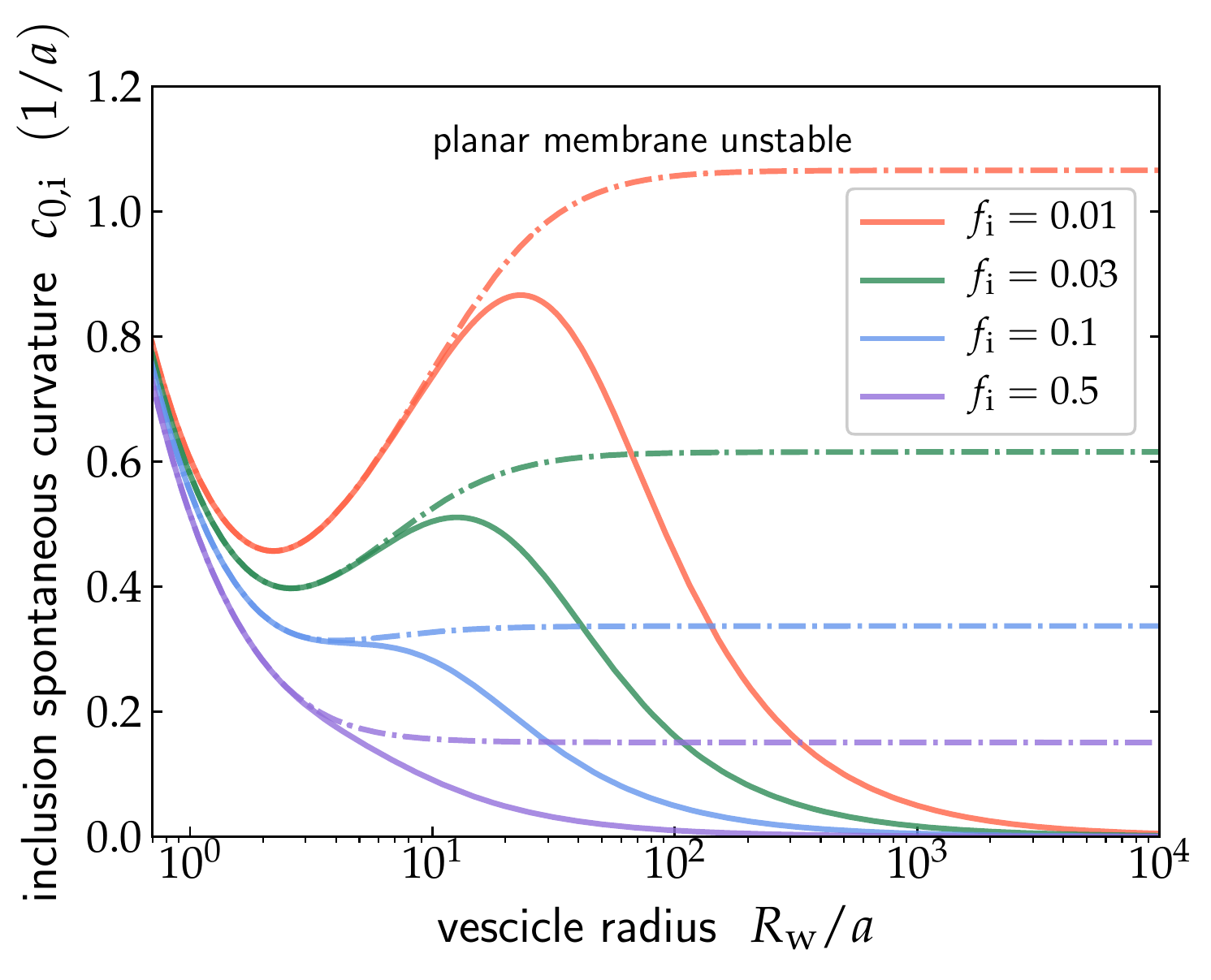}}
\caption{phase diagram of membrane stability. a) shows the linear and b) the logarithmic plot of the same data. Solid lines correspond to a membrane with a fraction $f_{\rm i}$ of inclusions with absolute spontaneous curvature $c_{\rm 0,i}$. Any point above the solid line thermodynamically unstable with respect to the vesicle formation and the flat membrane is only metastable. In the limit of large vesicles ($R_{\rm w} \to \infty $) the lines converge to zero ($c_{\rm 0,i} \to 0$) meaning that a flat membrane is always thermodynamically unstable. Dot-dashed lines show the corresponding phase plots where two types of inclusions are present such that the total spontaneous curvature of the membrane is zero: $c_{\rm 0,i'} = -c_{\rm 0,i}$ and $f_{\rm i'} = f_{\rm i}$. In this case the phase plots converge to a finite value of spontaneous curvature for large vesicles indicating that a flat membrane is thermodynamically stable. Parameters: $f_{\rm r}=0.0$, $\kappa = -\bar{\kappa} = 22 k_{\rm B}T$.}
\label{fig-memstab}
\end{figure*}

\subsection{Kinetics of endocytosis}
On Figures~\ref{fig-recr} and~\ref{fig-chRSI} we show additional simulation results of recruiting and endocytosis kinetics.  
%The effect of changing receptor interaction (Fig.~\ref{fig-chRSI}c)) has a similar effect on kinetics as introducing inclusions with spontaneous curvature (Fig.~\ref{fig-chRSI}b)) -- the rate is increases or decreased for all particle sizes $R_{\rm p}$. Introducing both positively and negatively curved inclusions (Fig.~\ref{fig-chRSI}a)) increases the endocytosis rate for smaller particles, but decreases it for larger particles. The scaling in the regime of large particles also deviates from the purely frictional  ($k_{\rm r}\propto 1/R_{\rm p}$) and begins approaching the diffusion limited regime ($k_{\rm r} \propto 1/R_{\rm p}^2$). We explain this behaviour by our finding that inclusions with positive spontaneous curvature occupy the "neck" region during endocytosis, as shown in Fig.~\ref{fig-neck}. Negatively curved inclusions and receptor must diffuse over the neck region to be able to fully envelop the particle. 

To further rationalise the enhanced sensitivity to cargo size when inclusions of both curvatures are present in the same amount, let us analyse the energetics of the neck of the membrane bud. The neck will be composed of two principal curvatures - the curvature parallel to the cargo-membrane contact line, and the curvature perpendicular to it. Following~\cite{agudo2015critical}, we define the effective adhesive length, $R_{\rm adh}$, which for the parameters used in Figure~\ref{fig-chR} is approximately $R_{\rm adh} = \sqrt{2\kappa / |W|} \approx 5 \sigma$, with $\kappa = 22k_{\rm B}T$ and the effective adhesion $|W| = \frac{\epsilon}{\sigma^2} \frac{f_{\rm r} e^{-\beta \epsilon}}{1 + f_{\rm r} e^{-\beta \epsilon}} \approx 2 k_{\rm B} T / \sigma^2$. The adhesive length $R_{\rm adh}$ determines the principal curvature perpendicular to the contact line in the neck $C_1=1/R_{\rm adh}$, while the principal curvature parallel to the contact line is determined by the particle size $C_2 = -1 /R_{\rm p}$. For large cargoes, when $R_{\rm p} \gg R_{\rm adh}$, the mean curvature in the neck is positive, which makes it populated by positive inclusions, hampering endocytosis. For cargoes that satisfy $R_{\rm p} \approx R_{\rm adh}$, the mean curvature in the neck vanishes, and the negatively curved inclusions can freely diffuse into the membrane area wrapped around the cargo, enhancing endocytosis. This analysis is supported by monitoring the neck region composition shown on Figure~\ref{fig-neck}. For large particles ($R_{\rm p}=32 \sigma$) the neck composition is strongly dominated by positive inclusions, while for small particles ($R_{\rm p}=5 \sigma$) the neck region has approximately equal fraction of both positive and negative inclusions. 
Interestingly, the composition of the wrapped shell shows opposite tendency: For large particles the fraction of inclusion types adsorb is approximately the same, while for small particles the adsorption of positive inclusions is suppressed. 

\begin{figure*}
\center
\subfigure[]{\includegraphics[width=0.32\textwidth]{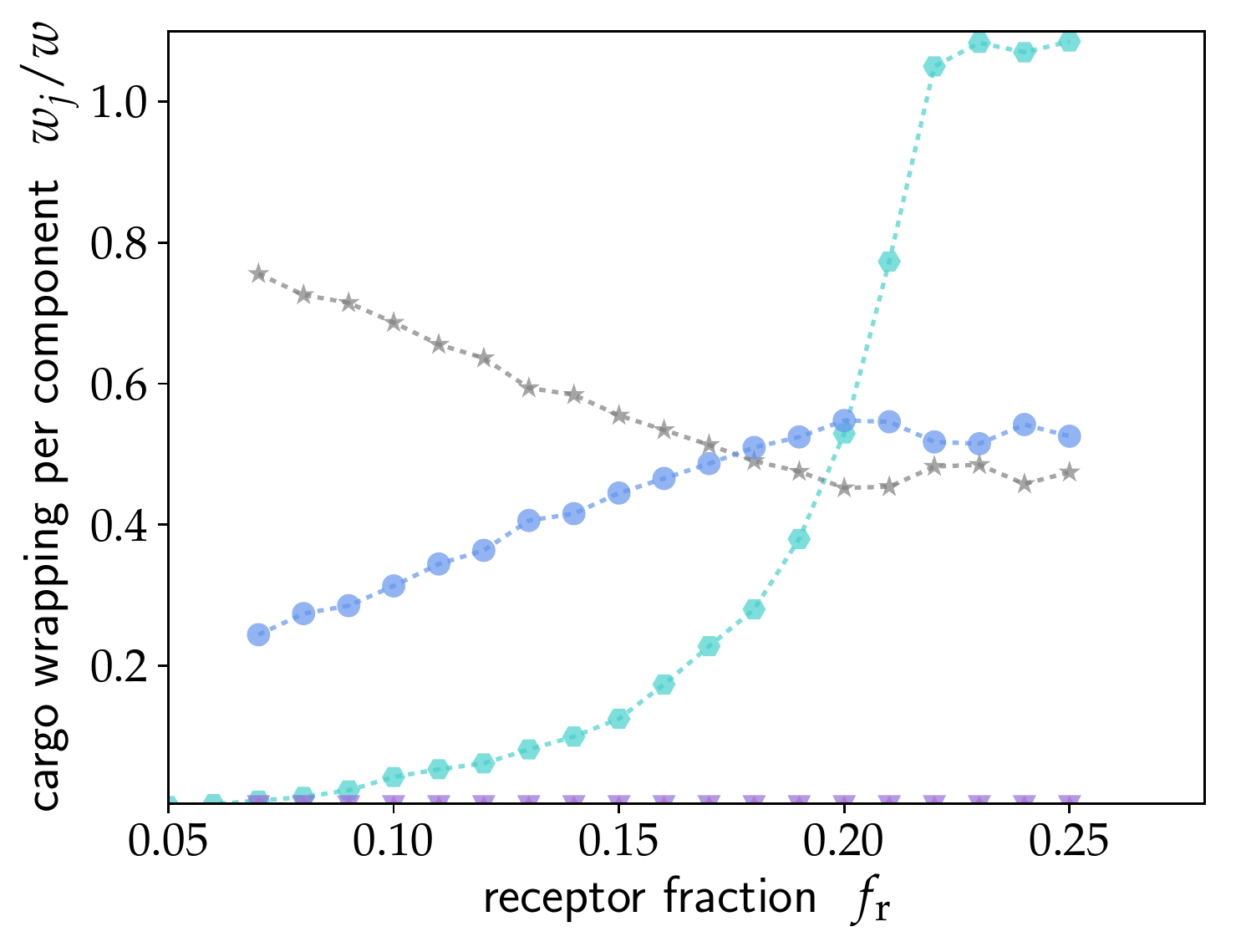}} \
\subfigure[]{\includegraphics[width=0.32\textwidth]{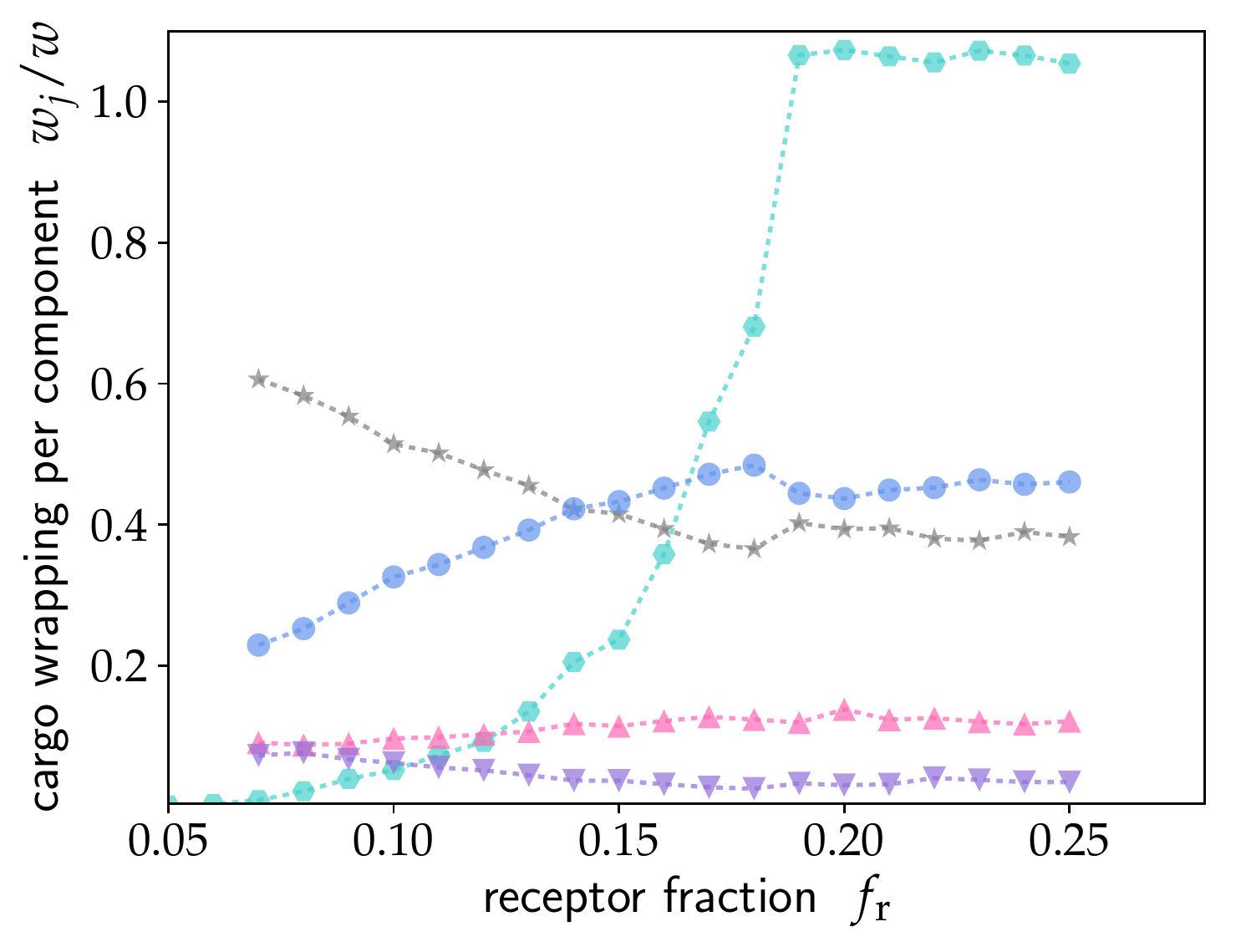}} \
\subfigure[]{\includegraphics[width=0.32\textwidth]{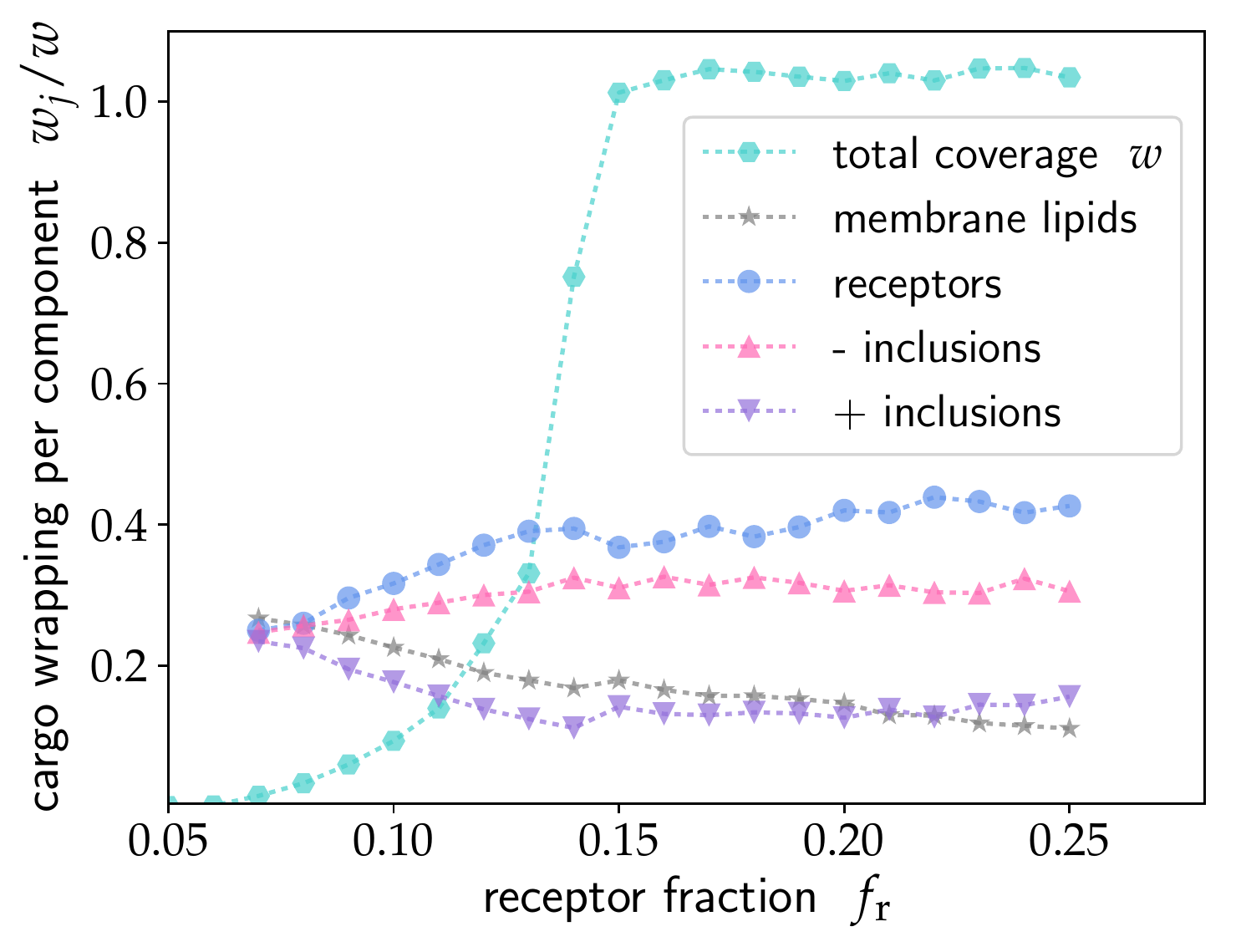}}
\caption{Recruiting of different types membrane inclusions. The turquoise hexagons on all three plots correspond to data on Figure~5A in the main text for a) $f_{\rm i}=f_{\rm i'}=0$, b) $f_{\rm i}=f_{\rm i'} = 0.1$ and c) $f_{\rm i} = f_{\rm i'}=0.3$. The grey stars, blue circles, pink and purple triangles show the fraction of beads of specific type in contact with the nanoparticle. Clearly, the receptor beads and negatively curved inclusions are recruited to the particle, while membrane beads and positively curved inclusions are expelled. Receptor fraction $f_{\rm r}=0.2$ and membrane bead fraction is $f_{\rm m} = 1-f_{\rm r} - 2f_{\rm i}$.}
\label{fig-recr}
\end{figure*}
%
%\begin{figure}
%\centering
%\subfigure[]{\includegraphics[width=0.47\textwidth]{Fig6pm_loglog.pdf}} \
%\subfigure[]{\includegraphics[width=0.47\textwidth]{Fig6kin_loglog.pdf}} \
%\subfigure[]{\includegraphics[width=0.47\textwidth]{Fig6eps_loglog.pdf}}  \
%\caption{\textbf{Endocytosis rate depends non-monotonically on the particle size.} a) Logarithmic plot of the data on Figure 6 in the main text. The black and red solid line indicate the power law scaling $k_{\rm r}\propto R_{\rm p}^{-\gamma}$ with the scaling exponent $\gamma$ = 1.1 and 1.4, respectively. b) Logarithmic plot of data in the inset of Figure 6 in the main text. c) Logarithmic plot endocytosis rate $k_{\rm r}$ at different receptor interaction strength $\epsilon_{\rm r}$ without present inclusions $f_{\rm i}=0$, receptor fraction is the same in all three plots: $f_{\rm r}=0.4$. }
%\label{fig-chRSI}
%\end{figure}

\begin{figure}
\centering
\includegraphics[width=0.5\textwidth]{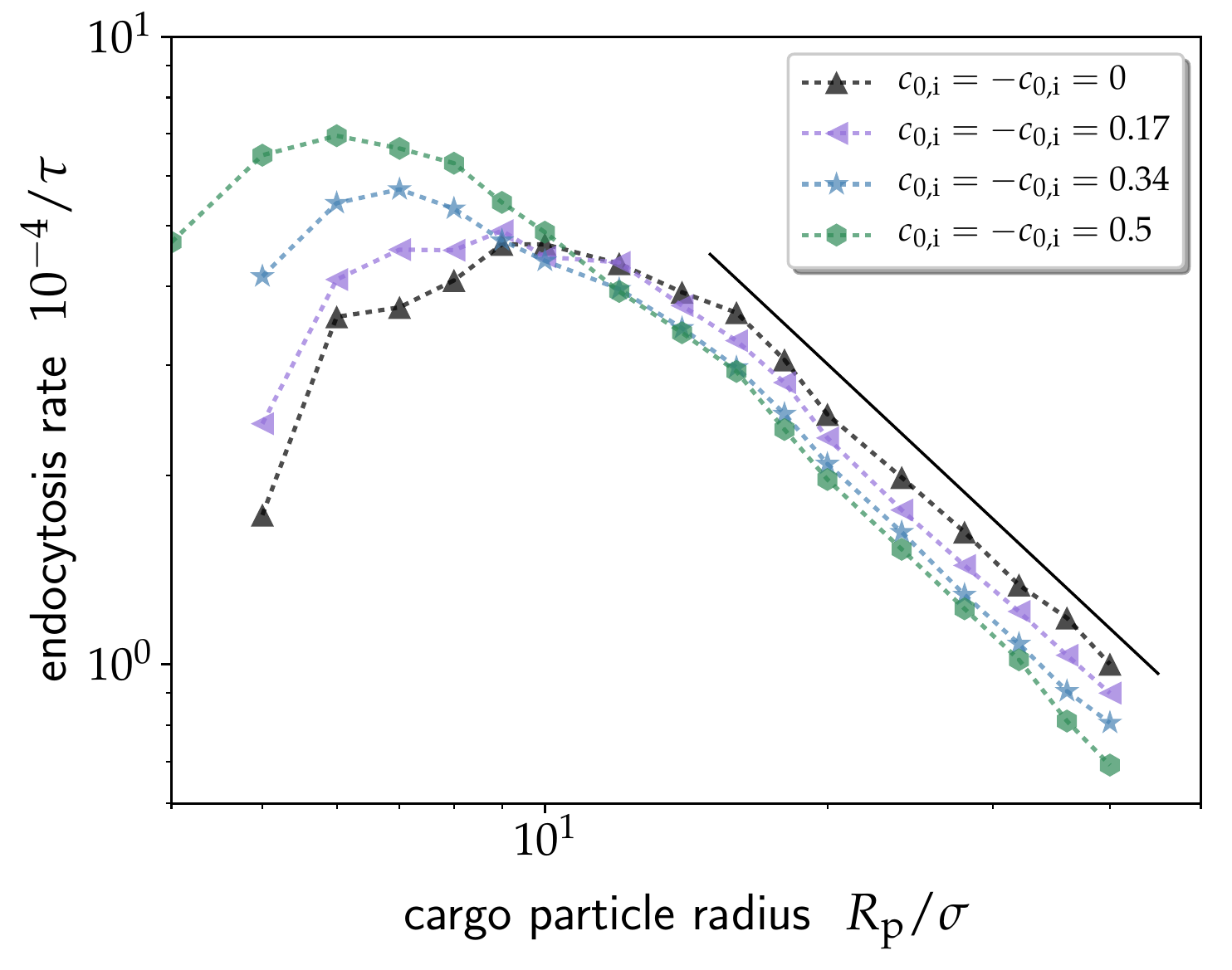} 
%\subfigure[]{\includegraphics[width=0.5\textwidth]{Fig6pm_loglog.pdf}} \
%\subfigure[]{\includegraphics[width=0.47\textwidth]{Fig6kin_loglog.pdf}} \
%\subfigure[]{\includegraphics[width=0.47\textwidth]{Fig6eps_loglog.pdf}}  \
\caption{\textbf{Endocytosis rate depends non-monotonically on the particle size.} a) Logarithmic plot of the data on Figure 5 in the main text. The black solid line indicates the power law scaling $k_{\rm r}\propto R_{\rm p}^{-\gamma}$ with the scaling exponent $\gamma$ = 1.4. }
\label{fig-chRSI}
\end{figure}

\begin{figure*}[ht!]
\center
\subfigure[\ large cargo,  $R_{\rm p}=32\sigma$]{\includegraphics[width=0.47\textwidth]{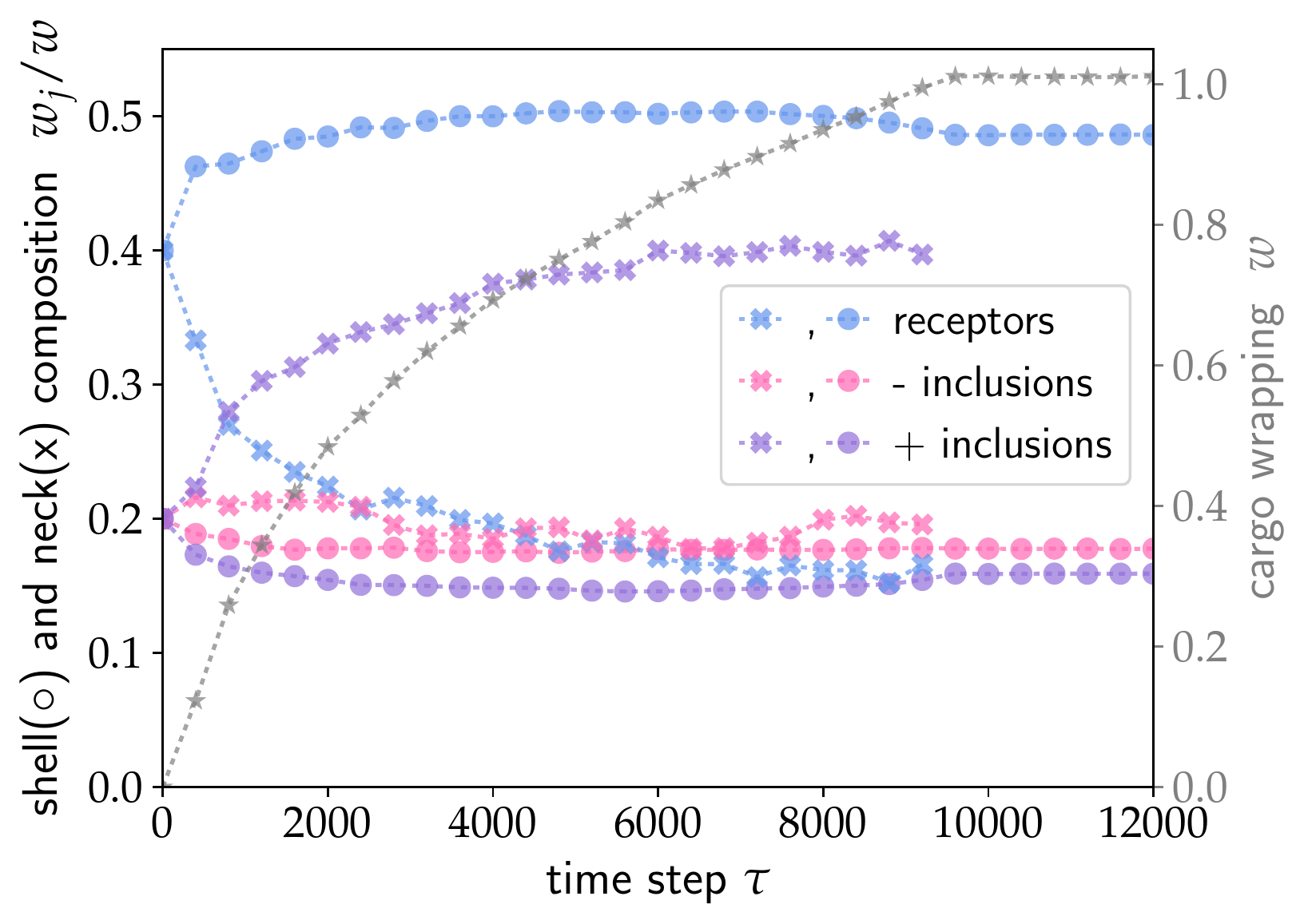}} \ \ \ \ \
\subfigure[\ small cargo, $R_{\rm p}=5\sigma$]{\includegraphics[width=0.47\textwidth]{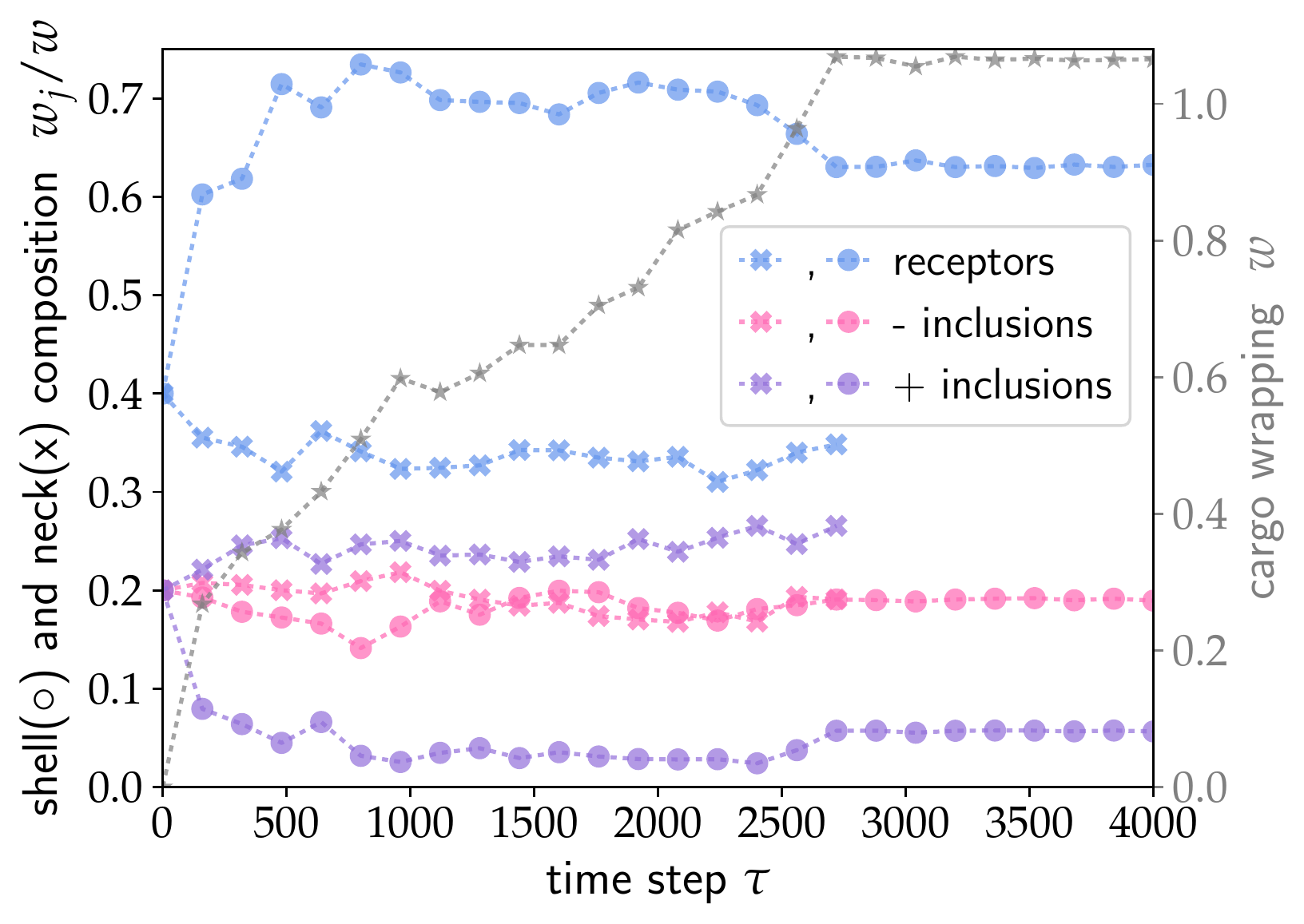}} \\
\subfigure[\ large cargo, $R_{\rm p}=5\sigma$]{\includegraphics[width=0.40\textwidth]{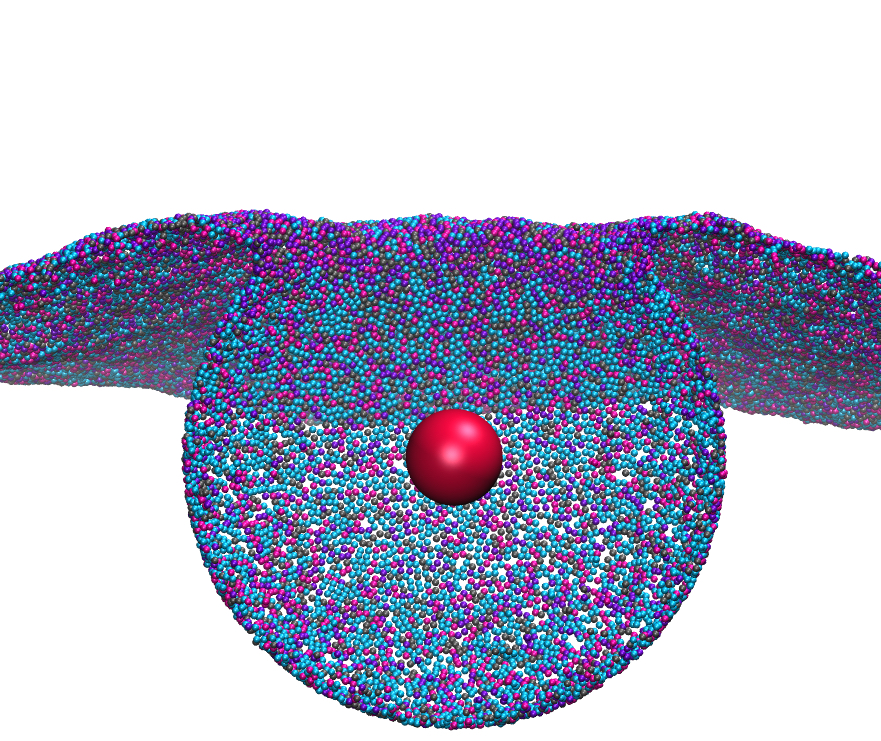}} \ \ \ \ \ \ \ \
\subfigure[\ small cargo, $R_{\rm p}=5\sigma$]{\includegraphics[width=0.40\textwidth]{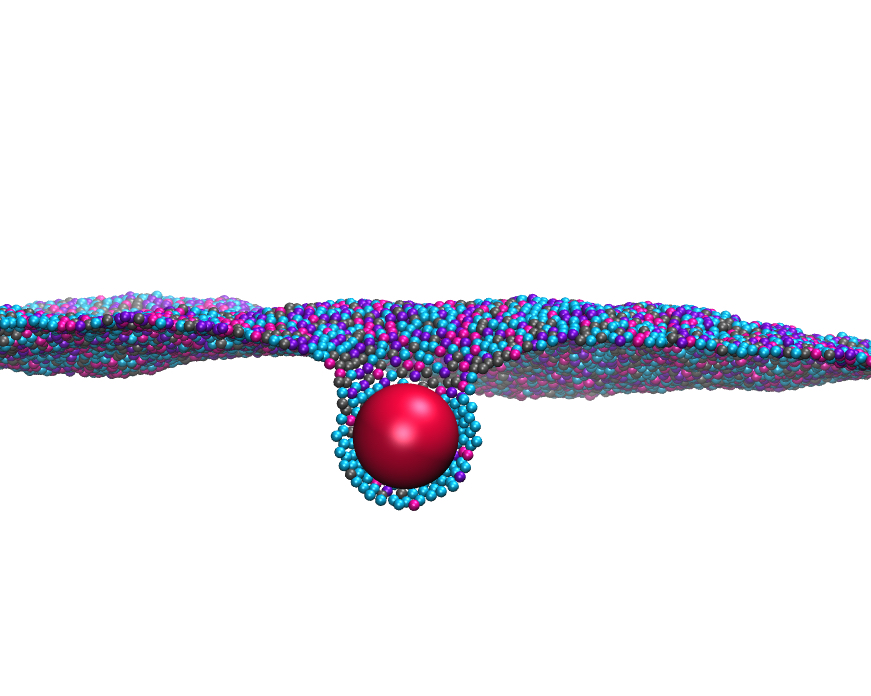}} 

\caption{\textbf{Analysis of the composition of the membrane neck as the endocytosis proceeds}. In a) and b) the circles denote the composition of the membrane shell wrapped around the cargo while the crosses show the composition of the membrane neck. Receptor beads are coloured blue, negative inclusions are pink color and positive inclusions purple. The cut-through snapshots parallel to the membrane plane on c) and d) correspond to plots on a  and b), respectively, at cargo wrapping of approximately $w=0.75$. The color scheme matches the symbol colors in a) and b), membrane (lipid) beads are colored grey. Note that the cargo size in c) has depicted using a significantly smaller radius for better visualisation of the wrapped membrane shell. The wrapped shell is defined as all beads within distance $1.5\sigma$ of the particle surface. The neck region is defined as all particles between a distance $1.5\sigma - 8\sigma$ of the particle surface. Grey stars show the total cargo wrapping (right axis) indicating the progression of endocytosis. Parameters correspond to Figure 5 in the main text: $f_{\rm r}=0.4$, interaction $\epsilon^*_{\rm r}=2.5k_{\rm B}T$ and curvature $c_{0,{\rm r}}= 0$. The inclusion fraction is $f_{\rm i} = f_{\rm i'}=0.2$ with opposite spontaneous curvatures $c_{0,{\rm i}}= - c_{0,{\rm i'}}= 0.34/\sigma$.}
\label{fig-neck}
\end{figure*}

\end{widetext}

\end{document}